\newif\ifnew\newtrue
\ifnew
     \documentclass[12pt]{article}
  \else
     \documentstyle[12pt]{article}
\fi
\newcommand{\eqn}[1]{(\ref{#1})}
\newcommand{\ft}[2]{{\textstyle\frac{#1}{#2}}}
\newsavebox{\uuunit}
\sbox{\uuunit}
    {\setlength{\unitlength}{0.825em}
     \begin{picture}(0.6,0.7)
        \thinlines
        \put(0,0){\line(1,0){0.5}}
        \put(0.15,0){\line(0,1){0.7}}
        \put(0.35,0){\line(0,1){0.8}}
       \multiput(0.3,0.8)(-0.04,-0.02){12}{\rule{0.5pt}{0.5pt}}
     \end {picture}}
\newcommand {\unity}{\mathord{\!\usebox{\uuunit}}}
\def\trace{{\rm Tr}\hskip 1pt}
\newcommand{\dsl}{\not\!\partial}
\newcommand{\dr}{\raise.3ex\hbox{$\stackrel{\leftarrow}{\partial }$}{}}
\newcommand{\delr}{\raise.3ex\hbox{$\stackrel{\leftarrow}{\delta }$}{}}

\csname @addtoreset\endcsname{equation}{section}
\newcommand{\dk}{\delta_\kappa }
\newcommand{\dkt}{\delta_\kappa \theta}
\begin{document}
\begin{titlepage}
\begin{flushright}
SU-ITP-97/48\\
KUL-TF-97/34\\
hep-th/9711161\\
November 20, 1997
\end{flushright}
\vspace{.5cm}
\begin{center}
\baselineskip=16pt
{\Large\bf M 5-brane and superconformal (0,2)\\
\
tensor multiplet in 6 dimensions}
\vskip 0.3cm
{\large {\sl }}
\vskip 10.mm
{\bf  Piet Claus$^+$,  ~Renata Kallosh$^{*}$ ~and
{}~Antoine Van Proeyen$^{*,+,\dagger}$  } \\
\vfill
{\small
$^*$
 Physics Department, \\ Stanford
University, Stanford, CA 94305-4060, USA
\\ \vspace{6pt}
 $^+$
Instituut voor theoretische fysica, \\
Katholieke Universiteit Leuven, B-3001 Leuven, Belgium
}
\end{center}
\vfill
\par
\begin{center}
{\bf ABSTRACT}
\end{center}
\begin{quote}
We present a gauge-fixed M 5-brane action: a 6-dimensional field
theory of a self-interacting (0,2) tensor multiplet with 32
worldvolume supersymmetries. The quadratic part of this action is
shown to be invariant under rigid $OSp(8^*|4)$ superconformal
symmetry, with 16 supersymmetries and 16 special supersymmetries.
\par
We explore a deep relation between the superconformal symmetry on
the worldvolume of the brane and symmetry of the near horizon anti-de Sitter
infinite throat geometry of the M 5-brane in space-time.
\vfill
 \hrule width 5.cm
\vskip 2.mm
{\small
\noindent $^\dagger$ Onderzoeksdirecteur FWO, Belgium }
\end{quote}
\end{titlepage}
\section{Introduction}
M 5-brane \cite{guven} is one of the most interesting supersymmetric extended
objects with the rare property of being completely non-singular \cite{GHT}.
The recently discovered kappa-symmetric action of the M 5-brane is the most
mysterious one
\cite{5b,ag}. The superembedding
approach to the M 5-brane theory was developed in \cite{HSW} and it
gives a covariant set of equations of motion of the theory in a beautiful
geometric setting.
The gauge-fixed M 5-brane action describes a 6-dimensional action of a
self-interacting (0,2) tensor multiplet with 32 worldvolume supersymmetries
\cite{RenataQuant}.
\par
There is a growing interest to a 6-dimensional superconformal theory
which should describe the small fluctuations (and possibly
interactions) of the M 5-brane (M 5-branes) \cite{strom,
wit,
seibsxt}. The standard lore here goes as follows: ``The $(0,2)$
theory of $k$ 5-branes of M theory has a moduli space of vacua
$(R^5)^k/S_k$. At the origin of the moduli space, the theory is
superconformal and has $U(k)$ gauge symmetry" \cite{BR}. It seems
also that not much of this is actually understood. The most
difficult part of making this picture realistic is related to our
inability at present to construct a non-Abelian interaction of $k$
(0,2) supersymmetric tensor multiplets. This is different from the
picture of interacting D-branes where the underlying Yang-Mills
field theory of non-Abelian vectors fields is available. In fact,
not much is known
even about the superconformal theory of one tensor multiplet in
d=6. Only the  on shell superconformal tensor multiplet in d=6 as well as
an on shell conformal supercurrent superfield are known
\cite{HST}.
\par
The purpose of this paper is to study various aspects of the
M 5-brane theory. We  present a detailed exposition and
derivation of the  $\kappa$-symmetric M 5-brane action
\cite{5b, ag} in our notation which are given in Appendix A. We further present
a detailed derivation of  the gauge-fixed  action \cite{RenataQuant}
describing the
self-interacting (0,2) tensor multiplet and its 32 global
supersymmetries, half of which are of the Volkov-Akulov type.
\par
The quadratic part of this action, the free action of the tensor
supermultiplet, will be explicitly shown to be invariant under
superconformal symmetry with 16 supersymmetries and 16 special
supersymmetries. This superconformal symmetry of the 6-dimensional
theory has a bosonic part with $SO(6,2)=SO(8^*)$-symmetry, which is exactly
the symmetry of the space-time M 5-brane near the horizon. The near
horizon geometry is given by $adS_7\times S^4$ \cite{GibTown}. The
superconformal symmetry forms an $OSp(8^*|4)$ algebra whose bosonic
part is $SO(6,2) \times USp(4)$.
\par
In the process of proving the superconformal symmetry of the (0,2)
free tensor multiplet action we have found that in general, except
in the very first paper on Wess-Zumino model \cite{WZ}, the concept
of rigid superconformal symmetry was not developed to the extent
which is required to establish the {\it presence or absence of
superconformal symmetry of generic non-gravitational theories}.
In \cite{ErginTanii} rigid superconformal transformations for
scalar multiplets in presence of a background metric are discussed.
In
most cases the {\it local superconformal symmetry} was developed in
the past, in particular in 6-dimensional case the superconformal
matter couplings to supergravity were presented in \cite{d6conf}.
Since the actions of kappa-symmetric branes gauge-fixed in a flat
target space are not gravitational actions but rather actions of
matter multiplets, we have worked out the concept of {\it rigid
superconformal symmetry} in general.
\par
We observe\footnote{We are grateful to J. Maldacena who attracted
our attention to this} the fascinating {\it relation between the
enhancement of supersymmetry near the M 5-brane horizon and
emergence of superconformal supersymmetry} for the small
fluctuations of the 5-brane. The M 5-brane configuration in the bulk
breaks 1/2 of the supersymmetry of 11-dimensional supergravity. The
brane interpolates between the Minkowski ${\bf M}^{11}$ space at
infinity, which is maximally supersymmetric, and $adS_7\times S^4$
near the horizon, which is also maximally supersymmetric
\cite{GibTown}. The symmetry of the supersymmetric $adS_7$
space-time is the symmetry of the superconformal field theory on the
brane. At present we do not have a clear explanation of the relation
between enhancement of supersymmetry in space-time and emergence of
special conformal symmetry on the brane. However we have found the
relation between the bosonic symmetries of the anti-de Sitter space
and conformal symmetry on the brane.
\par
In the past it was often conjectured that conformal supersingleton
field theories will provide the action for small fluctuations of the
superbranes. A discussion of this conjecture with
respect to near horizon geometry of various branes with extensive
list of references can be found in the paper of Gibbons and Townsend
\cite{GibTown}. We recall the singletons are the most fundamental
representations of the anti-de Sitter group $SO(p+1,2)$. In case of
$SO(6,2)$ they are called doubletons. The supersymmetric singleton
(doubleton) field theories live on the boundary of $adS_{p+2} $
space, given by $S^p\times S^1$. The singleton field theories have
some number of scalars and spinors and the action has a
superconformal symmetry. The doubleton supermultiplet forms an
ultra-short representation of the $OSp(8^*|4)$ superconformal
algebra and has 5 scalars, a chiral spinor (0,2) and an
antisymmetric tensor with self-dual field strength \cite{GunNWar}. The
equation of motions of the doubleton theory, which lives on
$S^5\times S^1$ boundary of the $adS_7$ space-time have a
superconformal symmetry \cite{NicSezTa}.
\par
We will find that the small fluctuations of the M 5-brane are indeed
given by the doubleton supermultiplet of $adS_{7}$ group. However,
the quadratic part of the M 5-brane, gauge-fixed in a flat
11-dimensional space, defines fields on the 6-dimensional Minkowski
space, it is superconformal invariant and different from the
superconformal doubleton field theory since the equations of motion
are different. After verifying the superconformal symmetry of the
small excitations of the M 5-brane we will find that the
self-interaction of the single M 5-brane in a flat 11-dimensional
background violates the superconformal symmetry of the quadratic
approximation.
\par
The paper is organized as follows. In Sec. 2 we introduce the
concept of rigid conformal symmetry of non-gravitational theories.
For theories with supersymmetry the conformal symmetry leads to the
appearance of the second special supersymmetry via the commutator of
usual supersymmetry with special conformal symmetry. We discuss the
superconformal algebra in general case and the one relevant to M
5-brane theory, an $OSp(8^*|4)$ algebra. Using a `triality' of $SO(6,
2)$ one can write down a manifestly symmetric superalgebra with
graded (fermionic) bosonic (anti) commutators. In Sec. 3 we present
the $OSp(8^*|4)$ superconformal algebra in more familiar form with
only 6-dimensional symmetry manifest: we have $SO(5,1)$ Lorentz
symmetry, translations, special conformal symmetry and dilatation,
i.e. 28 generators of $SO(6,2)$ bosonic conformal symmetries. Next,
there are 16 supersymmetries and 16 special supersymmetries and also
the generators of internal symmetry, an $USp(4)\approx {Spin(5)}$
group. We further recall the properties of the doubleton
representation of the $OSp(8^*|4)$ superconformal algebra
\cite{GunNWar} and the doubleton field theory \cite{NicSezTa}.
\par
In Sec. 4 we present a free
theory of the (0,2) tensor multiplet in d=6 and prove that it has a
superconformal symmetry.
We contrast our superconformal action for the small excitations of
the M 5-brane with the doubleton field theory \cite{NicSezTa}.
\par
Sec. 5 describes the M 5-brane action in flat 11-dimensional space
and gauge-fixing of local symmetries on the brane, reparametrization
and kappa, which leads to the theory of the self-interacting (0,2)
tensor multiplet. The quadratic part of this action is shown to
coincide with the action of the free theory of the (0,2) tensor
multiplet. The interaction terms are shown to break superconformal
symmetry of the free action.
\par
In Sec. 6 we explore the connection between superconformal symmetry
on the worldvolume and M 5-brane as a classical BPS solution of 11d
supergravity which near the horizon tends to $adS_{7}\times S^4$
geometry with enhancement of supersymmetry near the horizon. We
display the relation between the linearly realized $SO(6,2)$ part of
superconformal symmetry of the $adS_{7}$ space (considered as a
hypersurface in an 8-dimensional space) and non-linearly realized
superconformal symmetry $SO(6,2)$ on the 6-dimensional worldvolume.
In Conclusion we list our main new results. Appendix A contains the
notations. They are rather extensive as we have to deal with spinor
structures in $SO(10,1)$, $SO(5,1)$and $SO(6,2) $ theories. Appendix
B has some useful information on rigid
superconformal symmetry for non-gravitational theories and finally,
Appendix C has details on the derivation of $\kappa$-symmetry of the M
5-brane.
\section{Superconformal symmetry in \ifnew \\ \else \newline \fi
non-gravitational theories}
The superconformal group in the past was used as a tool for obtaining
supergravity actions with matter couplings invariant under local
super-Poincar\'{e}
transformations  in various dimensions \cite{BdWconf,Karpacz2,SalamSezgin}.
First the action was build
with superconformal symmetry and then the conformal symmetry was
gauge-fixed:
the remaining action provided the action of supergravity coupled to
all possible
matter multiplets.
\par
At present there is a growing  interest to superconformal
non-gravitational theories.
In principle, for this purpose one can use the available information
in the literature on  local
superconformal theories and decouple supergravity to be left with
superconformal theories of matter multiplets. However, in practice,
this is a
rather complicated route. Therefore we will set up here a
general procedure   to
define superconformal non-gravitational theories directly avoiding
introducing
and decoupling gravity. This we call rigid superconformal symmetry.
First we
consider the bosonic part of the superconformal symmetry and after
that the
enlarging with supersymmetry.
\subsection{Conformal transformations in dimension $d>2$} \label{ss:conftr}
The bosonic part of conformal transformation in dimension $d$ includes
the Lorentz transformation $M_{\mu\nu} $ of $SO(d-1,1)$, the translation
$P_\mu$, the special conformal transformation $K_{\mu}$ and the
dilatation $D$
\cite{conformal}. Here $\mu=0,1,\ldots ,d-1$. All these
transformations can
be nicely packaged in an algebra of the conformal group $SO(d,2)$ with
generators $M_{\hat\mu\hat\nu}=-M_{\hat\nu\hat\mu}$,
$\hat\mu=0,1,\ldots
,d+1.$  Those  include the Lorentz generators $M_{\mu\nu} $ of
the subgroup
$SO(d-1,1)$, whereas  translations , special conformal
transformations  and
dilatations  form the rest of the $SO(d,2)$ algebra as follows:
\begin{equation}
P_\mu= 2(M_{\mu d}+M_{\mu (d+1)})  ,~~
K_\mu= 2(M_{\mu(d+1) }-M_{\mu d}) ,~~  D=2 M_{(d+1)d} \ .
\label{PKDSO}
\end{equation}
Thus the conformal group is $SO(d,2)$ defined by the algebra
\begin{equation}
\left[M_{\hat\mu\hat\nu},M_{\hat\rho\hat\sigma}\right]
=\eta_{\hat\mu[\hat\rho}M_{\hat\sigma]\hat\nu}
-\eta_{\hat\nu[\hat\rho}M_{\hat\sigma]\hat\mu}\ ,  \label{algSO}
\end{equation}
where $\eta$ is the diagonal metric $(-++... +-)$.
Corresponding to those conformal generators
$\{P_\mu , M_{\mu\nu}, D, K_\mu\}$, with the
parameters $\{a^\mu ,\lambda_M^{\mu\nu},\lambda_D ,
\Lambda_K^\mu\},$ the
generator of the
infinitesimal conformal transformation is:
\begin{equation}
\delta_C= a^\mu  P_\mu + \lambda_M^{\mu\nu}M_{\mu\nu}+\lambda_D D +
\Lambda_K^\mu K_\mu   \ .
\end{equation}
In general, fields $\phi^i(x)$ of the $d$-dimensional theory have
the following
transformations under conformal group:
\begin{eqnarray}
\delta_C \phi^i(x)&=& \xi^\mu(x)\partial_\mu \phi^i(x)
+ \Lambda_M^{\mu\nu}(x)\, m_{\mu\nu}{}^i{}_{j}\phi^j(x)
\nonumber\\ &&+ w_i\,\Lambda_D(x)
\,\phi^i(x)+ \Lambda_K^\mu\left( k_\mu \phi\right) ^i(x)\ .
\label{deltaC}
\end{eqnarray}
To specify for each field $\phi^i$ its transformations under conformal
group one has to specify:
\par
i) transformations under the Lorentz group, encoded into the matrix
$(m_{\mu\nu})^i{}_j$. The properties of the  matrix $m_{\mu\nu}$ which
defines the Lorentz rotation can be found in appendix~\ref{app:conf}.
\par
ii) The Weyl weights, $w_i$.
\par
iii) Possible extra parts of the
special conformal transformations ($k_\mu \phi)^i$ (apart from those
connected to translations, rotations and dilatations as shown below
in \eqn{ximu} and \eqn{Lambdax}). Examples of such
transformations can be found e. g. in case of fields forming a
representation of the real $N=1$ supermultiplet in d=4 with Weyl weight
$w$: in particular, under
special conformal transformations the last component  field ${\cal D}$
transforms as $\delta {\cal D} = -2w  \Lambda_K^\mu D_{\mu} {\cal
C}$ where
${\cal C}$ is the first component of the real superfield
\cite{Karpacz1,Karpacz2}.
The Weyl weight of $(k_\mu \phi^i)$ should be
$w-1$, and $k_\mu$ are mutually commuting operators.
\par
We now can explain the various terms and why the transformations
are called  {\it  rigid}. The
parameters
$\{a^\mu ,\lambda_M^{\mu\nu},\lambda_D , \Lambda_K^\mu\}$ are all
$x$-independent
global, but
the transformation of the fields does depend on the coordinates $x$
of the
$d$-dimensional
space via the $x$-dependent translation $\xi^\mu(x)$,
\begin{equation}
\xi^\mu(x)=a^\mu +\lambda_M^{\mu\nu}x_\nu+\lambda_D x^\mu
+(x^2\Lambda_K^\mu-2x^\mu x\cdot \Lambda_K) \ , \label{ximu}
\end{equation}
$x$-dependent rotation $\Lambda_{M\,\mu\nu}(x)$ and $x$-dependent
dilatation $\Lambda_D(x)$:
\begin{eqnarray}
\Lambda_{M\,\mu\nu}(x)&=&\partial_{[\nu}\xi_{\mu]}=\lambda_{M\,\mu\nu}
-4x_{[\mu} \Lambda_{K\,\nu]} \ , \nonumber\\
\Lambda_D(x)&=&\ft1d \partial_\rho \xi^\rho =
\lambda_D -2 x\cdot \Lambda_K  \ .   \label{Lambdax}
\end{eqnarray}
Here the $x$-dependent translation $\xi^\mu(x)$ is the Killing vector
satisfying
\begin{equation}
\partial_{(\mu}\xi_{\nu)}-\ft1d \eta_{\mu\nu}\partial_\rho \xi^\rho=0\ .
\end{equation}
In d=2 the  Killing equations are reduced to
$\partial_z \bar \xi = \partial_{\bar z}  \xi = 0$
and this leads to an infinite dimensional conformal algebra.
In dimensions $d>2$ the conformal algebra is finite-dimensional.
\par
With these rules the conformal algebra is satisfied. The
question remains when an action is conformal invariant. We consider
local actions which can be written as $S=\int d^dx\,{\cal
L}(\phi^i(x),\partial_\mu\phi^i(x))$, i.e. with at most first order
derivatives on all the fields. For $P_\mu$ and $M_{\mu\nu}$ there are
the usual requirements of a covariant action. For the local
dilatations we have the requirement that the weights of all fields in
each term should add up to $d$, where $\partial_\mu$ counts also for
1, as can be seen from \eqn{delCderphi}. Indeed, the explicit $\Lambda_D$
transformations finally have to cancel with
\begin{equation}
\xi^\mu(x)\partial_\mu{\cal L}\approx -(\partial_\mu \xi^\mu (x)){\cal
L}=-d\Lambda_D(x){\cal L}\ .
\end{equation}
For special conformal transformations one remains then with
\begin{equation}
\delta_K S=2\Lambda_K^\mu  \int d^dx\,
\frac{{\cal L}\dr}{\partial(\partial_\nu\phi^i)}
\left(- \eta_{\mu\nu}w_i\phi^i +2m_{\mu\nu}{}^i{}_j\phi^j \right)
+\Lambda_K^\mu\frac{ S\delr}{\delta \phi^i(x)}(k_\mu\phi)^i(x)\ .
\label{deltaLK}
\end{equation}
where $\dr$ indicates a right derivative. The first terms originate
from the $K$-transformations contained in \eqn{ximu} and \eqn{Lambdax}.
In most cases these are sufficient to find the invariance and no
$(k_\mu\phi)$ are necessary. In fact, the latter are often excluded
because of the requirement that they should have Weyl weight $w_i-1$,
and in many cases there are no such fields available.
\par
Although we will show that this condition is satisfied for many
dilatational invariant theories, it is non-trivial. As a
counterexample we give the action of the scalars $\phi^1$ and
$\phi^2$ (with Weyl weights $(\ft d2-1)$)
\begin{equation}
{\cal L}=\left( 1+\frac{\phi^1}{\phi^2}\right)
(\partial_\mu\phi^1)  (\partial^\mu\phi^2)\ .
\end{equation}
 When proving special conformal symmetry of the (0,2) theory, we
 will use \eqn{deltaLK} describing in general case the variation of the
action under
special conformal symmetries.
\par
Thus the lesson from this section is that to establish a rigid conformal
invariance of a non-gravitational action in general one has to use the
transformations on fields  as given
above,  see also Appendix B for additional details. The
special conformal symmetry in general does not follow from
Poincar\'{e} and
dilatation symmetry and has to be established independently.
\subsection{The supersymmetric algebra}
Enlarging the conformal algebra with supersymmetry $Q$ one automatically
gets a
second `special' supersymmetry $S$ as the commutator of $K$ and $Q$.
Furthermore in the anticommutator of $Q$
and $S$ appears apart from the conformal transformations also an
`internal' symmetry group. The algebras were classified in \cite{Nahm}.
To satisfy the theorem of Haag, {\L}opusza\'nsky and Sohnius, \cite{HLS}
the conformal group should appear as a factored subgroup of the
bosonic part of the superalgebra, and the fermionic generators should
sit in a spinorial representation of that group. In 6 dimensions this is
satisfied with  $OSp(8^*|2N)$. At first sight, this would give spinor
generators in the fundamental of $SO(8^*)$. But $SO(6,2)$ has
`triality'. This means that the left-handed spinor, the right-handed
spinor and the
vector representations are all equivalent. We will therefore first
rotate the generators $M_{\hat \mu\hat \nu}$ to antisymmetric objects
in chiral (right handed) spinor space.
This is done using the invertibility of
$ \left( \hat \Gamma_{\hat \mu\hat \nu}\right) ^{\hat \alpha\hat
\beta}$:
\begin{equation}
\left( \hat \Gamma_{\hat \mu\hat \nu}\right) ^{\hat \alpha\hat \beta}
\left( \hat \Gamma^{\hat \rho\hat \sigma}\right) _{\hat \alpha\hat
\beta} =
16\delta_{[\hat \mu}^{[\hat \rho}
\delta_{\hat \nu]}^{\hat \sigma]}\ .
\end{equation}
See appendix~\ref{app:Cliff62} for the properties of gamma matrices in
$(6,2)$ space. These properties imply that $\hat {\cal C}
\hat \Gamma_{\hat \mu\hat \nu}$ are the only antisymmetric ones in
chiral space (as they form already 28 independent matrices).
We thus define
\begin{equation}
M_{\hat \alpha\hat \beta}=\frac{1}{4}
\left( \hat \Gamma^{\hat \mu\hat \nu}\right) _{\hat \alpha\hat \beta}
M_{\hat \mu\hat \nu} \ ;\qquad
M_{\hat \mu\hat \nu}=\frac{1}{4}
\left( \hat \Gamma_{\hat \mu\hat \nu}\right)^{\hat \alpha\hat \beta}
M_{\hat \alpha\hat \beta}    \ .
\end{equation}
In the new form, the $SO(6,2)$ algebra is
\begin{equation}
\left[M_{\hat\alpha\hat\beta},M_{\hat\gamma\hat\delta}\right] =
\hat {\cal C}_{\hat\alpha[\hat\gamma}M_{\hat\delta]\hat\beta}
-\hat {\cal C}_{\hat\beta[\hat\gamma}M_{\hat\delta]\hat\alpha}\ ,
\label{algSOsp}
\end{equation}
where the charge conjugation matrix (symmetric in $(6,2)$)
plays now the role of metric.
\par
For (0,2) theories we want 16
supersymmetry generators + 16 special supersymmetries, which we find
in $OSp(8^*|4)$, which implies that the internal symmetry group is
$USp(4)\approx \overline{SO(5)}$.
\par
Orthogonal groups are defined by the transformations between
vectors $V^{\hat\mu}$, leaving invariant the inner product $<V,V>\equiv
V^{\hat\mu}\eta_{\hat\mu\hat\nu}V^{\hat\nu}$ (symmetric metric $\eta$).
For orthosymplectic
groups we introduce superspace vectors $V^\Lambda=
(V^{\hat\alpha},V^i)$, where the first ones are bosonic (8 in our
case), and the latter ones fermionic (4 for us). The
group is then defined by the transformations $V^\Lambda\rightarrow
M^\Lambda{}_\Sigma V^\Sigma$, leaving invariant the inner product
\begin{equation}
<V,V>= V^\Lambda\eta_{\Lambda\Sigma}V^\Sigma\ ,
\end{equation}
where we introduced the orthosymplectic metric
\begin{equation}
\eta_{\Lambda\Sigma}=  \pmatrix{\hat {\cal C}_{\hat\alpha\hat\beta}&0\cr
0&\Omega_{ij}}\ ,
\end{equation}
graded symmetric, i.e. symmetric in the upper left part, and
the lower right part contains
 the antisymmetric metric of $USp(4)$.
\par
Raising and lowering indices with this metric, we can write the
generators with all indices down, and using the notation for signs
where $(-)^{\hat\alpha}=1$, and $(-)^i=-1$ (i.e. in these sign factors the
bosonic indices $\hat\alpha$ get the value 0, and $i$ is 1), we then have
\begin{equation}
M_{\Lambda\Sigma}=-M_{\Sigma\Lambda}(-)^{\Sigma\Lambda}\ ,
\end{equation}
which are of statistics $(-)^{\Sigma +\Lambda}$.
Thus there are (introducing for future reference the notation
$U_{ij}=U_{ji}$ for what will become the $USp(4)$ generators):
\begin{eqnarray}
\mbox{bosonic}&:& M_{\hat\alpha\hat\beta}=-M_{\hat\beta\hat\alpha}\ ,
\qquad U_{ij}\equiv M_{ij}=M_{ji} \ ,\nonumber\\
\mbox{fermionic}&:& M_{\hat\alpha i}=-M_{i\hat\alpha}\ .
\end{eqnarray}
All the (anti)commutators can
then be written by the rule
\begin{eqnarray}
\left[M_{\Lambda\Sigma},M_{\Gamma\Pi}\right\} &=&
\ft12 \eta_{\Sigma\Gamma}M_{\Lambda\Pi}+
\ft12 \eta_{\Lambda\Pi}M_{\Sigma\Gamma}(-)^{\Pi(\Gamma+\Sigma)}
\nonumber\\ &&-
\ft12 \eta_{\Sigma\Pi}M_{\Lambda\Gamma}(-)^{\Gamma\Pi}-
\ft12 \eta_{\Lambda\Gamma}M_{\Sigma\Pi}(-)^{\Sigma\Gamma}\ .
\label{superalgebra}\end{eqnarray}
The left hand side contains the graded commutator.
This rule is thus the same as for the orthogonal groups, see e.g.
\eqn{algSO}, apart from signs corresponding to the interchanges of
indices (and we used that the 2 indices on $\eta$ have the same
statistics).
\par
The bosonic subalgebra contains two parts. For all indices of the
bosonic type, we recover \eqn{algSOsp}. For the
bosonic operators with two fermionic indices
 we find the unitary symplectic algebra:
\begin{equation}
\left[U_{ij},U_{k\ell }\right]=
\Omega_{i(k}U_{\ell )j} +\Omega_{j(k}U_{\ell )i}\ .  \label{algUSp}
\end{equation}
\par
Due to the triality transformation, the commutators of
the conformal algebra generators and the supersymmetries are now
in the form
\begin{equation}
\left[M_{\hat \mu\hat \nu},M_{\hat \alpha i}\right]=-\ft14\left(
\hat \Gamma_{\hat \mu\hat \nu} \right)_{\hat \alpha }{}^{\hat \beta}
M_{\hat\beta i} \ ,
\end{equation}
while the symplectic transformations of the fermionic generators are
\begin{equation}
\left[ U_{ij},M_{k\hat \alpha}\right]=-\Omega_{k(i}M_{j)\hat \alpha}\ .
\label{commUf}
\end{equation}
The operator $D=2M_{76}$ in vector indices becomes now proportional
to $\Gamma_{76}$. According to \eqn{Gamma67} its eigenvectors
will thus distinguish the chiral and antichiral parts of a spinor in
6 dimensions.  This implies that the $Q$ and $S$ supersymmetries,
which have respectively weight $\frac{1}{2}$ and $-\frac{1}{2}$
under the dilatations, correspond with the parts in the
splitting of the spinors
$M_{i\hat \alpha}$ as in \eqn{chiral62spinor2}. We thus define
\begin{equation}
M_{i\hat \alpha} =\ft{1}{4}\pmatrix{Q_{\alpha'i}\cr S_{\alpha i}}\ .
\end{equation}
The anticommutators of these fermionic generators are
\begin{equation}
\left\{M_{\hat \alpha i},M_{\hat\beta j}\right\}=-\ft12
\left( \Omega_{ij}M_{\hat\alpha\hat\beta}
+\hat{\cal C}_{\hat\alpha\hat\beta}U_{ij}\right) \ ,
\end{equation}
which we will explicitize below.
\vspace{5mm}
 \par
Also here it is useful to introduce $x$-dependent supersymmetry
parameters  to describe rigid superconformal transformations
\cite{WZ}
\begin{eqnarray}
\epsilon(x)&=&\epsilon+\gamma_\mu x^\mu\eta\ , \label{epsilonx}
\end{eqnarray}
where $\epsilon$ and $\eta$ are the parameters of $Q$ and $S$
supersymmetry.
The full generator of the infinitesimal transformations of the
superconformal
group is given by
\begin{equation}
\delta_{SC}=\delta_C +\bar \epsilon Q +\bar \eta S +\delta_U\ ,\qquad
\delta_U=\ft12\alpha^{ij} U_{ij}=\alpha^{a'b'} U_{a'b'} \ ,
\end{equation}
where in addition to conformal transformations defined above we have
supersymmetry, special supersymmetry and internal symmetry
transformations~$U$.
\section{The superconformal algebra \ifnew \\ \else \newline \fi
in six dimensions }
Here we will rewrite the superalgebra $OSp(8^*|4)$ in \eqn{superalgebra},
relevant for six-dimensional (0,2)
non-gravitational theory, in more standard terms. First we consider the
28-parameter conformal group $SO(6,2)$ which include the 21
parameter Poincar\'e
group, dilatations and special
conformal transformations. The algebra is given by \footnote{In the
2-dimensional case the analogous set of generators ($M_{01}, D,
P_0, P_1, K_0,
K_1)$ corresponds to a finite subgroup of the infinite dimensional
conformal group and is well known in
terms of $L_{-1}= {1\over 2} (P_0 - P_1)$, $L_0={1\over 2}D+M_{10}$,
$L_1={1\over 2} (K_0 + K_1)$,
$\bar L_{-1}= {1\over 2} (P_0 + P_1)$, $\bar L_0={1\over 2}D-M_{10}$,
$\bar L_1={1\over 2} (K_0 - K_1)$. Higher order $L_n, |n| \geq 2 $ have
no analogs in $d > 2$. The $Q$ and $S$ supersymmetries are in $d=2$
the Neveu-Schwarz generators $G_{-\frac{1}{2}}$ and $G_{\frac{1}{2}}$
respectively.} \eqn{algSO}, which, using the definitions
\eqn{PKDSO}, gets the form
\begin{eqnarray}
[M_{\mu\nu} , M^{\rho\sigma}]&=&
-2\delta_{[\mu}^{[\rho} M_{\nu]}{}^{\sigma]} \ , \nonumber\\
\ [P_{\mu} , M_{\nu\rho}\ ] &=&\eta_{\mu[\nu} P_{\rho]}\ , \nonumber\\
\ [K_{\mu} , M_{\nu\rho}\ ] &=& \eta_{\mu[\nu}K_{\rho]} \ , \nonumber\\
\ [P_{\mu} , K_{\nu}\ ]&=& 2 (\eta_{\mu\nu} D + 2 M_{\mu\nu}) \ ,
\nonumber\\
\ [D , P_{\mu} \ ]&=& P_\mu \ , \nonumber\\
\ [D , K_{\mu} \ ]&=&-K_\mu \ .
\end{eqnarray}
On the fields this is realized as
\begin{equation}
\left[\delta_C(\xi_1),\delta_C(\xi_2)\right]=\delta_C\left(
\xi^\mu=\xi_2^\nu\partial_\nu\xi_1^\mu-\xi_1^\nu\partial_\nu\xi_2^\mu\right)\ .
\end{equation}
\par
The anticommutator of supersymmetries $Q$ produces translation,
that of the special
supersymmetries $S$ produces a special conformal transformation and
finally the
anticommutator of supersymmetry with the special supersymmetry
produces a
dilatation, a Lorentz
transformation and an $USp(4)\approx {Spin(5)}$ rotation $U^{ij}$:
\begin{eqnarray}
\{ Q_{\alpha'}^i , Q_{\beta'}^j\}&=& -2 (\gamma_\mu)_{\alpha' \beta'}
\Omega^{ij} P^\mu \ , \nonumber\\
\{ S_{\alpha'}^i , S_{\beta}^j\}&=& -2  (\gamma_\mu)_{\alpha \beta}
\Omega^{ij} K^\mu  \ , \nonumber\\
\{ Q_{\alpha'}^i , S_{\beta}^j\}&=& -2 c_{\alpha' \beta} (  \Omega^{ij}
D +4\, U^{ij})-   2 (\gamma_{\mu\nu})_{\alpha' \beta}
\Omega^{ij} M^{\mu\nu} \ .         \label{QSalgebra}
\end{eqnarray}
The $USp(4)$ part of the algebra is  \eqn{algUSp}.
There is also a set of commutators between generators of conformal
group and
fermionic generators which carry the information that the
supersymmetry and
special supersymmetry transform as spinors under the Lorentz group and
that the Weyl weight of $Q (S) $ equals $1/2 (-1/2)$ :
\begin{eqnarray}
\ [ M_{\mu\nu} , Q_{\alpha'}^i \ ] &=& -{1\over 4}
(\gamma_{\mu\nu}Q)_{\alpha'} ^i   \
, \qquad \ [ M_{\mu\nu} , S_{\alpha}^i \ ]= -{1\over 4}
(\gamma_{\mu\nu}S)_{\alpha} ^i \ ,
\nonumber\\
\ [ D , Q_{\alpha'}^i \ ]&=& {1\over 2}Q_{\alpha'} ^i   \ , \hskip
2.4 cm  \ [
D , S_{\alpha}^i \ ]= -{1\over 2} S_{\alpha} ^i \ .
\end{eqnarray}
There are two  important commutators: i) between special conformal
symmetry and
supersymmetry, which produces special supersymmetry. This means
that each time
when the theory has special conformal symmetry and supersymmetry,
the special
supersymmetry is
guaranteed;  ii) Translation and  special supersymmetry generate
supersymmetry.
\begin{equation}
\ [ K_{\mu} , Q_{\alpha'}^i \ ]=  (\gamma_{\mu}S)_{\alpha} ^i   \ ,
\hskip 2.0 cm
 \ [ P_{\mu} , S_{\alpha}^i \ ]=  (\gamma_{\mu}Q)_{\alpha'} ^i \ .
\end{equation}
Finally, there are commutators of supersymmetry and special
supersymmetry with
internal symmetry carry the information that the supersymmetry and
special
supersymmetry are in the fundamental representation of  the
internal group $USp(4)$ as given in \eqn{commUf}.
\par
This is a full set of generators of $OSp(8^*/4)$
supergroup which
is a super extension of the anti-de Sitter group in 7 dimensions.
All
positive-energy unitary representations of this supergroup are
constructed in
\cite{GunNWar} using super-oscillators. In particular, the ultrashort
multiplet,
called doubleton  was found in \cite{GunNWar} by using one pair of
oscillators.
The corresponding 6-dimensional field theory was expected to contain a
two-index antisymmetric tensor field with a self-dual field strength, 5
scalars
and a (0,2) chiral spinor. The superconformal doubleton field theory was
constructed in \cite{NicSezTa}.  In
fact due to the self-duality of the tensor fields only the field
equations were found.
The theory was constructed for the fields which live on the
boundary of the $adS_7$ which is an $S^p\times S^1$ space. The
equations of
motion were shown to transform
into each other under the action of the superconformal
transformations. In
particular, the equation of motion for the 5 scalars of the
doubleton field
theory was found to be
\begin{equation}
(\nabla^\mu \partial_\mu - 4) \phi^{a'} =0 \ ,
\label{doubleton}\end{equation}
where the covariant derivative is due to the fact that the fields
are coupled
to the curved background of $S^p\times S^1$. We will show below
that the
action of the free tensor multiplet in Minkowski 6-dimensional space is
superconformal invariant and
the equation of motion for 5 scalars, being a free equation,  is
different from \eqn{doubleton}.
\section{Small excitations of the M 5-brane }
We consider the following action which will be later shown to be a
quadratic
approximation of the gauge-fixed M 5-brane action:
\begin{equation}
S_{lin}=\int d^6x\, \left[-\ft12 H^-_{\mu\nu}H^{*\mu\nu}
- \ft12 \partial_\mu X^{a'}\cdot \partial^\mu X^{a'}
+2\bar \lambda\dsl\lambda\right]\ .
\end{equation}
This is the action of the free superconformal (0,2) tensor multiplet in  a
6-dimensional Minkowski
space \footnote{Recently an action of a free (0,1) tensor multiplet
was presented in \cite{Padova}. This action was shown to be
supersymmetric. Our proof that the action of (0,2) tensor multiplet
is not only supersymmetric but also superconformally symmetric can
be easily applied to (0,1) theory since our theory consists of a (0,
1) tensor multiplet and a hypermultiplet.}. There are 5
scalars
$X^{a'}$, a  right-handed  16-component spinor $\lambda$ and a
tensor field $B_{\mu\nu}$ with  on shell self-dual  field strength.  The
auxiliary field $a(x)$ was introduced by
Pasti-Sorokin-Tonin \cite {auxil}  in order to write down a 6-dimensional
Lorentz covariant action for a self-dual
tensor.
The derivative of the auxiliary field
$ v_\mu = {\partial_ \mu a \over \sqrt {(\partial_ \nu a)^2}}$ is used to
convert a 3-index field $H_{\mu\nu\rho}$ into 2-index field $H_{\mu\nu}$:
\begin{equation}
H_{\mu\nu\rho}=3\partial_{[\mu} B_{\nu\rho ]} \ , \qquad H_{\mu\nu}
\equiv
H_{\mu\nu\rho} v^\rho \ . \end{equation}
 The action depends on
\begin{equation}
H^*_{\mu\nu} = H^* _{\mu\nu\rho} v^\rho \ , \qquad H^*
_{\mu\nu\rho}\equiv
{1\over 6} \epsilon_ {\mu\nu\rho\sigma\tau\phi} H^{\sigma\tau\phi} \ ,
\end{equation}
and
\begin{equation}
H^-_ {\mu\nu} \equiv {1\over 2}( H_{\mu\nu} - H^*_{ \mu\nu}) \ .
\end{equation}
The action is invariant under the local symmetries I, II and III
discovered in
\cite{Padova}:
\begin{eqnarray}
&I)& \delta_I B_{\mu\nu}=2\partial_{[\mu}\Lambda_{\nu]}\ , \qquad \delta_I
a=0 \ ,\nonumber\\
&II)&
\delta_{II} B_{\mu\nu}=2 H_{\mu\nu}^{- }\frac{1}{\sqrt{u^2}}
\varphi\ , \qquad
\delta_{II}  a=\varphi  \ , \nonumber\\
&III)& \delta_{III} B_{\mu\nu}=\psi_{[\mu} v_{\nu]}\ , \qquad \delta_{III}
a=0\ .
\label{gauge}\end{eqnarray}
Using these  symmetries one can show that the field equation for
the tensor
field can be reduced to  the self-duality condition:
\begin{equation}
H^-_ {\mu\nu\rho} \equiv {1\over 2}( H_{\mu\nu\rho} - H^*_{ \mu\nu\rho})
=0\ .
\end{equation}
The transformations I and III are reducible gauge invariances.
First there is the usual one for gauge antisymmetric tensors.
But there is more:
\begin{eqnarray}
&a)& \Lambda_\mu=\partial_\mu \Lambda \ ,\nonumber\\
&b)& \psi_\mu=v_\mu \psi \ , \nonumber\\
&c)& \Lambda_\mu=u_\mu\Lambda' \ ,\qquad
\psi_\mu=-\sqrt{u^2}\partial_\mu\Lambda'\ .
\label{zeroIIII}
\end{eqnarray}
\par
The action is invariant under {\it rigid conformal symmetry} under
condition
that
we  assign the following weights to our fields
\begin{equation}
w (X^{a'})= w (B_{\mu\nu})=2 \ , \qquad w (a) =0 \ , \qquad w
(\lambda)={5\over
2} \ .
\end{equation}
It follows that
\begin{equation}
w (v_\mu)= 0  \ , \qquad w (H_{\mu\nu\rho})=w (H_{\mu\nu\rho} v^\rho)=3 \ .
\end{equation}
One can easily see the Poincar\'e symmetries.
For the special conformal transformations we have to check
\eqn{deltaLK}.
It is satisfied separately for $\phi^i$ representing the fields $a$,
$X^{a'}$, $\lambda$ and $B_{\mu\nu}$. In fact the mechanisms at work
are typical for the usual way in which dilatational invariant
theories are often conformal invariant:
\begin{enumerate}
\item $a$ is a scalar of Weyl weight 0, such that both terms in the
bracket of \eqn{deltaLK} are vanishing.
\item The term with $\phi^i\rightarrow X^{a'}$ is a total derivative.
\item $\lambda^i$ falls in the category of spinors  whose derivative
appears in the action as $\dsl\lambda$, and have Weyl weight
$(d-1)/2$. The various terms in \eqn{deltaLK} cancel.
\item \label{exampleconf} Similarly $B_{\mu\nu}$ with Weyl weight 2
belongs to the category of vectors or antisymmetric tensors whose
derivatives appear as
$ \partial_{[\mu_1}B_{\mu_2\ldots \mu_p]}$ with Weyl weight
$p-1$. Again with this weight the terms cancel in \eqn{deltaLK}.
This value of the Weyl weight is what we need also in order
that their gauge invariances and their zero modes commute with the
dilatations.
\end{enumerate}
\par
The action is invariant under the superconformal supersymmetry
transformations
with
left-handed parameter $\epsilon$ and right-handed $\eta$:
\begin{eqnarray}
\delta_\epsilon X^{a'}&=&-2\bar \epsilon(x)\gamma^{a'}\lambda \ ,\nonumber\\
\delta_\epsilon\lambda&=& \ft12(\dsl X^{a'})
\gamma_{a'}\epsilon(x)-\ft16
h^+_{\mu\nu\rho}\gamma^{\mu\nu\rho}\epsilon(x)+2X^{a'}\gamma_{a'}
\eta \ ,\nonumber\\
\delta_\epsilon B_{\mu\nu}&=&-2\bar \epsilon(x)\gamma_{\mu\nu}\lambda\ .
\end{eqnarray}
We remind that $\epsilon(x)=\epsilon+\gamma_\mu x^\mu\eta$ as
explained in
\eqn{epsilonx} with respect to rigid supersymmetries. Here the off shell
(anti)
self-dual tensor field strength is defined as
\begin{eqnarray}
h_{\mu\nu\rho}^\pm&\equiv&\ft14 H_{\mu\nu\rho}
-\ft32 v_{[\mu}H^\mp_{\nu\rho]}\nonumber\\
&=& v^\sigma\left( v_{[\sigma} H_{\mu\nu\rho]}\pm \ft16
\epsilon_{\mu\nu\rho}{}^{\tau\phi\chi}
v_{[\sigma}H_{\tau\phi\chi]}\right)\nonumber\\
&=&\pm\ft14 H^*_{\mu\nu\rho}-2v^\sigma v_{[\alpha}
H^\mp_{\beta\gamma\delta]}\ .
\label{defh}\end{eqnarray}
To prove the invariance one can use that
under variations of $a$ and $H_{\mu\nu\rho}$ we have
\begin{equation}
\delta S_{lin}=  \int d^6x\,
\left(\frac1{12}H^*_{\mu\nu\rho}-\frac23 h^{+}_{\mu\nu\rho}
\right)\delta H^{\mu\nu\rho} +\frac1{2\sqrt{u^2}}
\epsilon^{\mu\nu\rho\sigma\tau\phi} H^-_{\mu\nu}
H^-_{\rho\sigma}v_\tau\partial_\phi\delta a   \ .
\label{deltaIpm}
\end{equation}
 After taking the $\epsilon$-variations one adds
partial derivatives such that the derivatives do not act on
$\lambda$. Then one finds that there only remain derivatives on
$\epsilon$ from the variation of $\lambda$. We thus get
\begin{equation}
\delta_\epsilon S_{lin}=2\int d^6x\, \bar \lambda \gamma^\mu
\left((\dsl X^{a'})\gamma_{a'}-\ft13
h^+_{\nu\rho\sigma}\gamma^{\nu\rho\sigma}\right)
\partial_\mu\epsilon(x)\ .
\end{equation}
Using $\partial_\mu\epsilon(x) =\gamma_\mu\eta$, the second term drops
due to the identity $\gamma^\mu \gamma^{\nu\rho\sigma} \gamma_\mu =0$.
The first term can then be cancelled by the explicit $\eta$ term in
$\delta\lambda$.
Thus we have established that the action has an off shell superconformal
supersymmetry as well as a set of gauge symmetries defined in
\eqn{gauge}.
\par
The equations of motion for the scalars of the free superconformal
tensor
multiplet in Minkowski space are
\begin{equation}
\partial^\mu \partial_\mu  X^{a'} =0 \ ,
\label{scalars}\end{equation}
which is different from the equations for the doubletons
\eqn{doubleton} which was presented in Sec. 3.
\par
One can calculate the algebra of such transformations. The easiest
part is the
action of supersymmetries and special supersymmetries on the 5 scalars
$X^{a'}$. It is given by
\begin{eqnarray}
\left[\delta(\epsilon_1(x)),\delta(\epsilon_2(x))\right]X^{a'}&=&
\delta_C\left(\xi^\mu=2\bar
\epsilon_1(x)\gamma^\mu\epsilon_2(x)\right)X^{a'}\label{algXap}  \\ &&
+\delta_U\left(\alpha^{a'b'}= 4
\bar \epsilon_1\gamma^{a'b'}\eta_2 - 4
\bar \epsilon_2\gamma^{a'b'}\eta_1\right) X_{a'}\nonumber\ .
\end{eqnarray}
This expression is a convenient way of representing all the
anticommutators in \eqn{QSalgebra}.
Note that the expression for $\xi^\mu$ implies
\begin{eqnarray}
&&a^\mu= 2\bar \epsilon_1\gamma^\mu\epsilon_2\ , \qquad
\lambda_M^{\mu\nu}=2\bar \epsilon_1\gamma^{\mu\nu}\eta_2
-2\bar \epsilon_2\gamma^{\mu\nu}\eta_1\ ,\nonumber\\
&&\lambda_D= 2\bar \epsilon_1\eta_2 -2\bar \epsilon_2\eta_1  \ , \qquad
\Lambda_K^\mu = 2\bar \eta_1\gamma^\mu\eta_2 \ .
\end{eqnarray}
The auxiliary field $a$ is a singlet on all the fermionic symmetries
\cite{BergKal}. The algebra is
\begin{equation}
[\delta_1 , \delta_2] a = 2\bar\epsilon_1 \gamma^\mu \epsilon_2
\partial_\mu a +\phi(x) =0 \ ,
\end{equation}
which defines the gauge II transformation parameter $\phi(x) = -
2\bar\epsilon_1
\gamma^\mu \epsilon_2 \partial_\mu a (x)$ to be a function of a
derivative of
the $a$-field. On $B_{\mu\nu}$ field the algebra will be most
complicated, as
it will include all
3 type of gauge transformations I, II, III depending on fields and
bilinear in
susy and special susy parameters. On fermions the algebra will
contain terms
with equations of motion. Altogether this is an example of a so
called `open' and `soft gauge algebra'
\cite{soft}, which often appear in supersymmetric gauge theories,
supergravity and conformal theories. `Open algebra' refers to the
fact that equations of motion are necessary to close the algebra, the
algebra is exact on physical states. `Soft algebra' means that
the commutators close {\it with field-dependent structure functions}:
\begin{equation}
[L_i , L_j] =  c_{ij}^k L_k \ ,
\end{equation}
$\partial_\mu c_{ij}^k\neq 0$. Thus
the `structure constants' of the group are not really constant
but are field dependent and therefore space-time dependent.
Starting from this
type of a soft gauge algebra which follows from the symmetries of a given
Lagrangian, one can
define a so called `hard gauge algebra' where all fields are taken to be
constant (consistent with the field equations). All structure
constants of
the algebra become then indeed constants $\partial_\mu c_{ij}^k=0$. This
can be understood as  a bona fide algebra.
 It does not have to be associated with any
particular Lagrangian and could have a potential of describing
theories with
symmetries of a given supergroup even in absence of a particular
field theory Lagrangian. The
corresponding abstract algebra for the 6-dimensional superconformal
group was presented in the previous section on pure algebraic basis.
After identification of the `soft algebra' from the dynamical
Lagrangian of the tensor multiplet, one can verify that the `hard
algebra' derived from this Lagrangian is exactly the one given in Sec. 3.
\section{Classical action  of the M 5-brane.}
The classical kappa-symmetric action of the M 5-brane was discovered
by Bandos, Lechner, Nurmagambetov, Pasti, Sorokin and Tonin in
Lorentz covariant form \cite{5b} and by Aganagic, Park, Popescu and
Schwarz \cite{ag} in Lorentz-non-covariant form. The covariance was achieved
in \cite{5b} with the help of an auxiliary scalar field and
additional gauge symmetries (type II and III). Upon gauge-fixing of
these additional gauge symmetries a covariant M 5-brane action can
be reduced to a non-covariant form \cite{ag}.
\par
The form of both actions is rather complicated, the Wess-Zumino term
was not yet derived in an explicit form and finally there is a
technical problem associated with different notations, sometimes not
fully presented, in all available M 5-brane papers.
\par
Therefore we will first present a detailed form of the Lorentz
covariant classical action in a flat 11-dimensional background
together with the complete set of notations in Appendix~A and more
details on $\kappa$-symmetry in Appendix~C. We will also give an
explicit form of the WZ terms.
\subsection{The action and its symmetries}
We start from the following fields on the 6-dimensional world-volume of
the 5-brane.
\begin{equation}\label{fields}
 X^M(x)\ ;\qquad \theta(x)\ ;\qquad B_{\mu\nu}(x)\ ;\qquad a(x)\ .
\end{equation}
Here 11 $X^M(x)$ and 32 $ \theta(x)$ are coordinates of the
11-dimensional superspace which however are not considered as labels
only but as dynamical variables on the brane. $B_{\mu\nu}(x)$ is the
tensor field on the brane and $a(x)$ is an auxiliary scalar.
The action \cite{5b} in a flat 11-dimensional background is
\begin{equation}
S_{M5}=\int d^6 x \left(- \sqrt{-\det (g_{\mu\nu}+i\,  {\cal H}^*{}_{\mu\nu})}
-\frac{\sqrt{g}}4 {\cal H}^{*\mu\nu}
{\cal H}_{\mu\nu} \right)
+\int_{{\cal M}_7} I_7\ , \label{SM5}
\end{equation}
The first term in the action is given as an integral over the
6-dimensional volume and the second term, Wess-Zumino term, is
presented in the form of an integral over the 7-dimensional manifold
${\cal M}_7$, which has as a boundary the 6-dimensional
worldsheet, and $I_7$ is a total derivative (see section~\ref{ss:WZ}).
We used the following notations:
\begin{eqnarray}
&& g_{\mu\nu}=\Pi^M_\mu \eta_{MN}\Pi^N_\nu\ ; \qquad g=-\det g_{\mu\nu}
\nonumber \\
&&\Pi^M_\mu= \partial_\mu X^M -\bar \theta \Gamma^M \partial_\mu\theta
\nonumber \\
&& H=dB\ ,\qquad \mbox{which is }H_{\mu\nu\rho}=3\partial_{[\mu}
B_{\nu\rho ]}\nonumber \\
&& {\cal H}=H-  c_3\ ,\qquad \mbox{which is } {\cal H}_{\mu\nu\rho}=
3\partial_{[\mu} B_{\nu\rho ]}-  c_{\mu\nu\rho}\nonumber \\
&& u_\mu=\partial_\mu a\ ;\qquad u^2=u_\mu g^{\mu\nu}u_\nu\nonumber \\
&& v_\mu=\frac{u_\mu}{\sqrt{u^2}}  \ ;\qquad v^2=1\nonumber \\
&& {\cal H}_{\mu\nu}=v^\rho{\cal H}_{\mu\nu\rho}\nonumber \\
&& {\cal H}^*_{\mu\nu}=v^\rho{\cal H}^*_{\mu\nu\rho}
\ ,\qquad\mbox{where }{\cal H}^*_{\mu\nu\rho}=\frac{\sqrt{g}}{6}
\epsilon_{\mu\nu\rho\sigma\tau\phi} {\cal H}^{\sigma\tau\phi}
\nonumber \\
&& {\cal H}^\pm_{\mu\nu}=v^\rho{\cal H}^\pm_{\mu\nu\rho}\nonumber \\
&& I_7=R_7-\ft1{2}  {\cal H} R_4\nonumber\\
&& R_4=\ft12d\bar \theta\Gamma_{MN}d\theta \Pi^M \Pi^N=dc_3 \nonumber \\
&& R_7= dc_6 -\ft 12 c_3 \wedge R_4=\frac 1{5!}\Pi^{M_1} ... \Pi^{M_5}
d\bar\theta \Gamma_{M_1 ... M_5}d\theta\ .
\label{defn}
\end{eqnarray}
Here $\mu,\nu$ indices are raised or lowered with $g_{\mu\nu}$.
The expression $R_4$ is $d$-closed, which shows the consistency of the
definition of $c_3$, while the consistency of the definition of $c_6$
follows from
\begin{equation}
2dR_7 +R_4R_4 =0\ . \label{dR7}
\end{equation}
Explicit forms of $c_3$ and $c_6$ will be given below.
It is further useful to introduce a notation
\begin{equation}
{\cal G}= \sqrt{-\det (g_{\mu\nu}+i\,  {\cal H}^*{}_{\mu\nu})}\ .
\end{equation}
\par
The gauge symmetries of the classical action in 6 dimensions are:
\begin{itemize}
\item local diffeomorphisms
\item $\kappa$-symmetry (infinitely reducible)
\item I, II, III   (reducible)
\begin{eqnarray}
&I)& \delta_I B_{\mu\nu}=2\partial_{[\mu}\Lambda_{\nu]}\ ;\qquad
\delta_I
a=0\nonumber\\
&II)&
\delta_{II} B_{\mu\nu}=\frac{1}{\sqrt{u^2}} \varphi
\left( {\cal H}_{\mu\nu}+2\frac{\delta{\cal G}}
{\partial {\cal H}^{*\mu\nu}}\right) \ ;\qquad
\delta_{II}  a=\varphi   \nonumber\\
&III)& \delta_{III} B_{\mu\nu}=\psi_{[\mu} v_{\nu]}\ ;\qquad \delta_{III} a=0\
{}.
\end{eqnarray}
For the zero modes of I, and III, see \eqn{zeroIIII}.
\end{itemize}
Furthermore there are rigid symmetries in 11 dimensions
\begin{itemize}
\item Poincar\'e (translations with parameter $a^M$ and rotations
$\Lambda_{MN}$).
\begin{equation}
\delta_P X^M= -a^M -\Lambda^{MN}X_N\ ;\qquad
\delta_P \theta= -\ft14\Gamma_{MN}\Lambda^{MN} \theta
\end{equation}
For later convenience we took a minus sign for the translations in 11
dimensions.
\item supersymmetry
\end{itemize}
Under rigid space-time supersymmetry and 6-dimensional local
$\kappa$-symmetry
\begin{eqnarray}\label{transformations}
&&\delta_\epsilon \theta=\epsilon\ ;\qquad \delta_\epsilon X^M
=\bar \epsilon\Gamma^M\theta
\nonumber\\
&&\delta_\kappa\theta=(1+\Gamma)\kappa\ ;\qquad \delta_\kappa X^M
=\bar \theta\Gamma^M
\delta_\kappa\theta\ .
\end{eqnarray}
The auxiliary field $a$ does not transform under
$\epsilon$-supersymmetry and $\kappa$-transformations. The matrix
$\Gamma$ and the
$\kappa$-transformation of $B_{\mu\nu}$ will be given in
section~\ref{ss:kappagf}. For giving its $\epsilon$-transformation,
we first write
\begin{equation}\label{R4}
\delta_\epsilon R_4= 0\ ;\qquad R_4=dc_3\ ; \qquad \delta_\epsilon c_3=
d c_2(\epsilon)\ .
\end{equation}
This leads to explicit expressions \cite{deAzc}
\begin{eqnarray}\label{c3}
c_3&=&\ft12 \bar\theta \Gamma_{MN} d\theta\left( \Pi^M\Pi^N+\bar\theta
\Gamma^M d\theta
\Pi^N+\ft13 \bar\theta\Gamma^M d\theta\bar\theta\Gamma^N d\theta \right)
\nonumber\\
c_2(\epsilon)&=&\ft12 \bar\epsilon \Gamma_{MN} \theta \left(
\Pi^M\Pi^N +\ft53
\bar\theta\Gamma^M d\theta \Pi^N +\ft{11}{15} \bar\theta\Gamma^M
d\theta
\bar\theta\Gamma^N d\theta \right)\nonumber\\ &&
+\ft12 \bar \epsilon \Gamma^M\theta\, \bar\theta \Gamma_{MN} d\theta
\,\left( -\ft13 \Pi^N
-\ft4{15} \bar \theta\Gamma^N d\theta \right)\ .
\end{eqnarray}
We can thus choose the transformation of $B_{\mu\nu}$ such that
${\cal H}$ is invariant under rigid supersymmetry:
\begin{equation}
\delta_\epsilon {\cal H}=0\ \rightarrow\  \delta_\epsilon B=
c_2(\epsilon)\ .
\end{equation}
\subsection{The Wess-Zumino term}\label{ss:WZ}
Now we will define the Wess-Zumino term.  Due to \eqn{dR7} we  have
that $dI_7=0$, and \cite{CandielloLechner}
\begin{eqnarray}\label{I7}
I_7=dc_6-\ft12 H R_4= d\left( c_6-\ft12 B R_4 \right)=
d\left( c_6+\ft12 H c_3 \right)\ .
\end{eqnarray}
The Wess-Zumino term can thus also be defined as the 6-dimensional
integral of one
of the expression in brackets in the equation above. Its
normalization can be fixed by the requirement that the invariance
III should be valid\footnote{Alternatively, but more difficult, it is
fixed by $\kappa$-symmetry.}. We have that
\begin{eqnarray}
\delta_{III} H_{\mu\nu\rho}&=& 3\sqrt{u^2}v_{[\mu}\partial_\nu
\psi_{\rho]}\frac{1}{\sqrt{u^2}}\ ;\qquad
\delta_{III}{\cal H}^*_{\mu\nu}=0\label{deltaIIIHH}\\
\delta_{III} \int d^6 x \sqrt{g}{\cal H}^{*\mu\nu} {\cal H}_{\mu\nu}&=&
-\ft13\int d^6 x \sqrt{g}c^{*\mu\nu\rho}\delta_{III}H_{\mu\nu\rho}
=2 \int (\delta_{III} H) c_3 \ ,    \nonumber
\end{eqnarray}
such that the appropriate normalization is
\begin{eqnarray}
I_{WZ}&=&\int_{{\cal M}_7} I_7  =\int_{{\cal M}_6}
\left( c_6+\ft12 H c_3 \right)\nonumber\\
&=&\int d^6x\,\frac{1}{6!}\epsilon^{\mu _1\ldots \mu _6} \left(
c_{\mu _1\ldots \mu _6}-10 H_{\mu _1\mu _2\mu _3}c_{\mu _4\mu _5\mu _6}
\right)\ .
\end{eqnarray}
\par
To get an explicit form for $c_6$, using the above result for $c_3$,
we have to solve
\begin{eqnarray}
d c_6 &=& \ft 1{5!} d \bar \theta \Gamma_{M_1\dots M_5} d\theta
\Pi^{M_1}\dots
\Pi^{M_5}\nonumber\\
&&+\ft18 \bar \theta \Gamma_{M_1 M_2} d\theta d\bar \theta \Gamma_{M_3
M_4}d \theta
(\Pi^{M_1} \dots \Pi^{M_4} + \bar \theta \Gamma^{M_1} d \theta
\Pi^{M_2} \dots
\Pi^{M_4}\nonumber\\
&&\hskip4.5cm + \ft 13 \bar \theta \Gamma^{M_1} d \theta \bar \theta
\Gamma^{M_2}
d \theta \Pi^{M_3} \Pi^{M_4})   \ .
\end{eqnarray}
We find the following solution (determined up to
total derivatives)
\begin{eqnarray}
c_6 &=& \bar \theta \Gamma_{M_1\cdots M_5} d \theta
\big ( \ft1{5!} \Pi^{M_1} \cdots \Pi^{M_5} +
\ft 1{48} \bar \theta \Gamma^{M_1} d \theta \Pi^{M_2} \cdots \Pi^{M_5}
\nonumber\\
&&\hskip2.2cm + \ft 1{36} \bar \theta \Gamma^{M_1} d \theta
\bar \theta \Gamma^{M_2} d\theta \Pi^{M_3}\Pi^{M_4}\Pi^{M_5}\nonumber\\
&&\hskip2.2cm + \ft1{48} \bar \theta \Gamma^{M_1}d \theta
\bar \theta \Gamma^{M_2}d \theta\bar \theta \Gamma^{M_3}d \theta
\Pi^{M_4}
\Pi^{M_5}\nonumber\\
&&\hskip2.2cm + \ft 1{5!} \bar \theta \Gamma^{M_1}d \theta
\bar \theta \Gamma^{M_2}d \theta \bar \theta \Gamma^{M_3}d \theta
\bar \theta \Gamma^{M_4}d \theta \Pi^{M_5}\nonumber\\
&&\hskip2.2cm + \ft1{6!}  \bar \theta \Gamma^{M_1}d \theta
\bar \theta \Gamma^{M_2}d \theta\bar \theta \Gamma^{M_3}d \theta
\bar \theta \Gamma^{M_4}d \theta\bar \theta \Gamma^{M_5}d \theta
\big)\nonumber\\
&&-\bar \theta \Gamma_{M_1 M_2} d\theta \bar \theta \Gamma_{M_3
M_4} d \theta \big(\ft1{24} \bar \theta \Gamma^{M_1} d \theta
\Pi^{M_2}\Pi^{M_3}\Pi^{M4}\nonumber\\
&&\hskip4cm +\ft1{48} \bar \theta \Gamma^{M_1} d \theta
\bar \theta \Gamma^{M_2} d\theta \Pi^{M_3}\Pi^{M4}\nonumber\\
&&\hskip4cm +\ft1{5!} \bar \theta \Gamma^{M_1} d \theta
\bar \theta \Gamma^{M_2} d \theta \bar \theta \Gamma^{M_3} d \theta
\Pi^{M4} \big)\ .\label{c6}
\end{eqnarray}
\par
With indices we thus have
\begin{eqnarray}
c_{\mu\nu\rho}&=&-3\bar \theta
\gamma_{[\mu\nu}\partial_{\rho]}\theta+\ldots
\nonumber\\
c_{\mu_1\ldots \mu_6}&=&6 \bar \theta \gamma_{[\mu_1\ldots \mu_5}
\partial_{\mu_6]}\theta   +\ldots \ .
\end{eqnarray}
Here we have introduced the notation
\begin{equation}
 \gamma_\mu=\Pi_\mu^M \Gamma_M\ ; \qquad \gamma_{\mu\nu}=\gamma_{[\mu}
\gamma_{\nu]}\ ,\ldots\ .
\end{equation}
This accomplishes the derivation of the explicit form of the M
5-brane action.
\par
We still have to show that
the M 5-brane action upon gauge-fixing in a flat 11-dimensional
background  leads to a quadratic action for the tensor multiplet
which in the quadratic approximation
was shown to have a superconformal symmetry.
\subsection{$\kappa$-transformations and gauge fixing}
\label{ss:kappagf}
The form of the $\kappa$-transformations acting on all fields is
defined by the form of $\kappa$-transformations acting on $\theta$:
\begin{equation}
\delta_\kappa \theta (x) =(1+ \Gamma)\kappa (x) \ . \label{dktrule}
\end{equation}
Here $\Gamma$ is a function of fields of the theory:
\begin{eqnarray}
\Gamma &=&  \frac{1}{{\cal G}}
\left( \sqrt{g} \bar \gamma + \ft{\sqrt{g}}2{\cal H}_{\mu\nu}^*v_\rho
\gamma^{\mu\nu\rho} - \ft1{16}\epsilon^{\mu\nu\rho\sigma\tau\phi}
{\cal H}^*_{\mu\nu} {\cal H}^*_{\rho\sigma} \gamma_{\tau\phi}\right)
\nonumber\\
 \bar \gamma&=& \frac 1{6! \sqrt{g} }\epsilon^{\mu_1 ... \mu_6}
\gamma_{\mu_1 ... \mu_6}\ ,       \label{resGamma}
\end{eqnarray}
satisfying $\Gamma^2=1$.
On the remaining fields the $\kappa$-transformations are
\begin{eqnarray}
\delta_\kappa B &=& \ft12 \bar\theta \Gamma_{MN} \delta_\kappa
\theta \big(
\Pi^M\Pi^N + \bar\theta\Gamma^Md\theta\Pi^N - \ft13
\bar\theta\Gamma^Md\theta \bar\theta\Gamma^Nd\theta\big) \nonumber\\
&& - \ft12 \bar\theta \Gamma^M \delta_\kappa \theta \bar \theta
\Gamma_{MN}d
\theta\big( \Pi^N + \ft23 \bar\theta \Gamma^N d\theta\big) \nonumber\\
\delta_\kappa \Pi^M&=&2d\bar \theta\Gamma^M\dkt \ ,
\qquad \delta_\kappa a(x) =0 \,.   \label{deltakapparem}
\end{eqnarray}
We will summarize a proof of these results in appendix~\ref{app:kappa}.
\par
The $\kappa$-symmetry defined above is infinite reducible since it is
given in terms of 32-component spinor $\kappa$, which however one can
change  by  $(1- \Gamma)\kappa_1 (x)$, for any function $\kappa_1$ since
$(1+ \Gamma)(1- \Gamma)\kappa_1 (x)=0$.
\par
We first define the irreducible $\kappa$-symmetries, following
\cite{RenataQuant}. The matrix $\Gamma$ in \eqn{resGamma} has square 1,
such that the transformations where $\kappa$ is proportional to
$1-\Gamma$ are zero modes of \eqn{transformations}. To identify the
irreducible part, we consider first the classical values. We consider
the classical solution of the equations of motion
\begin{equation}
X^\mu=x^\mu\ ;\qquad \theta_\alpha=\theta_{\alpha'}=0\ ;\qquad
X^{a'}=\mbox{constant}\ ;\qquad B_{\mu\nu}=0\ .
\end{equation}
Then
\begin{equation}
\left.\Gamma\right|_{cl}=
\left.\bar\gamma\right|_{cl}= \Gamma_* \ .
\end{equation}
\par
Therefore the irreducible $\kappa$ symmetries are $\kappa_\alpha$,
and we put
\begin{equation}
\ft12(1-\Gamma_*)\kappa=\pmatrix{0\cr \kappa_{\alpha'}}=0\ .
\end{equation}
\par
The tracelessness and unipotency of $\Gamma$ allow to write it in
terms of $16\times 16$ matrices as
\begin{equation}
\Gamma=\pmatrix{C&B\cr A&-C'}\ \mbox{with }\trace C'=\trace C
\mbox{ and }AC=C'A\ ,
\end{equation}
and if $A$ is invertible: $B=(1-C^2)A^{-1}=A^{-1}(1-C'^2)$.
With the reduced $\kappa$-symmetries, we thus have under the
fermionic symmetries \eqn{transformations}
\begin{eqnarray}
\delta_f\theta_\alpha&=&\epsilon_\alpha+(1+C)_\alpha{}^\beta\kappa_\beta
\nonumber\\
\delta_f\theta_{\alpha'}&=&
\epsilon_{\alpha'}+A_{\alpha'} {}^\beta \kappa_\beta\ .
\end{eqnarray}
In the classical limit $A=0$ and $C=1$, therefore one can take the
gauge
\begin{equation}
\kappa\mbox{ - gauge : }
\ft12(1+\Gamma_*)\theta=\pmatrix{\theta_\alpha\cr 0}=0\ .
\end{equation}
The corresponding decomposition law is\footnote{For convenience,
we indicate here $L$ for left-handed and $R$ for right-handed spinors.}
\begin{equation}
\delta \theta_L=0\ \Rightarrow\ \kappa_L=(1+C)^{-1}
\left(-\epsilon_L +\ft12\gamma_{a'}\gamma_a
\Lambda^{a'a}\lambda_R\right) \ ,
\end{equation}
where we have renamed the remaining fermion is $\theta_{\alpha'} $, as
$\lambda$, which is then right-handed.
\par
We also gauge-fix the 6-dimensional diffeomorphisms by
\begin{equation}
\mbox{6-dimensional diffeomorphisms - gauge : }X^a(x)=\delta^a_\mu x^\mu\ .
\end{equation}
The resulting decomposition law is :
\begin{eqnarray}
\delta X^\mu&=&0\qquad\Rightarrow \nonumber\\
&&\delta^a_\mu\xi^\mu=
a^a +\Lambda^a{}_b \delta^b_\nu x^\nu +\Lambda^a{}_{a'}X^{a'}-
\bar \epsilon \gamma^a \theta
-\bar \lambda_R\gamma^a A\kappa_L\ . \label{decompxi}
\end{eqnarray}
This implies that after the gauge fixing the fields $X^{a'}$ and
$\lambda$ transform under the 32 rigid supersymmetries as follows
\begin{eqnarray}
\delta_\epsilon X^{a'}&=&-2\bar \epsilon_L\gamma^{a'}\lambda_R
+\left( \partial_\mu X^{a'}  \right) \delta^\mu_a\bar \lambda_R
\gamma^a\left( \epsilon_R+A(1+C)^{-1}\epsilon_L\right) \nonumber\\
\delta_\epsilon\lambda_R&=&\epsilon_R-\left( A(1+C)^{-1}\right)
\epsilon_L\nonumber\\
&&+ \left( \partial_\mu \lambda_R \right) \delta^\mu_a\bar \lambda_R
\gamma^a\left( \epsilon_R+A(1+C)^{-1}\epsilon_L\right) \ .
\label{32susy}\end{eqnarray}
\par
We further have
\begin{equation}
\Pi_\mu^a=\delta_\mu^a-\bar \lambda\gamma^a\partial_\mu\lambda\ ;
\qquad
\Pi_\mu^{a'}=\partial_\mu X^{a'}\ . \label{Pigf}
\end{equation}
Due to the chirality restriction of the gauge fixed $\theta$, which
becomes now $\lambda$, there are further simplifications.
E.g.
\begin{equation}
c_{\mu\nu\rho}=3\bar \lambda\gamma_a\gamma_{a'}\partial_{[\mu}\lambda
\left( 2\delta_{\nu}^a-\bar \lambda\gamma^a\partial_\nu\lambda\right)
\partial_{\rho]}X^{a'}\ .
\end{equation}
\par
After gauge-fixing, we can use in the M 5-brane action \eqn{SM5} (action
of the self-interacting tensor multiplet)
\begin{equation}
g_{\mu\nu} = \eta_{\mu\nu} -\bar \lambda \gamma_a
(\delta^a_\mu \partial_\nu + \delta^a_\nu \partial_\mu) \lambda + \partial_\mu
X^{a'} \partial_\nu X^{a'} +
(\bar \lambda \gamma^a\partial_\mu \lambda) (\bar \lambda \gamma_a\partial_\nu
\lambda)
\label{metric}\end{equation}
and
\begin{equation}
{\cal H}_{\mu\nu\rho}= 3\partial_{[\mu} B_{\nu\rho ]}-  3\bar
\lambda\gamma_a\gamma_{a'}\partial_{[\mu}\lambda
\left( 2\delta_{\nu}^a-\bar \lambda\gamma^a\partial_\nu\lambda\right)
\partial_{\rho]}X^{a'}\ .
\label{3form}\end{equation}
The following explicit form
of the Born-Infeld expression ${\cal G}$
for what concerns the dependence on ${\cal H}^*_{\mu\nu}$ can be
given:
\begin{eqnarray} {\cal G}^2&=&
-\det(g_{\mu\nu} + i {\cal H}_{\mu\nu}^*)\label{BI}\\
&=& g \left(1 - \ft 12 {\cal
H}_{\mu\nu}^* {\cal H}^{*\mu\nu} + \ft18 ({\cal H}_{\mu\nu}^*{\cal
H}^{*\mu\nu})^2 -
\ft14({\cal H}_{\mu\nu}^*{\cal H}^{*\nu\rho}{\cal H}_{\rho\sigma}^*{\cal
H}^{*\sigma\mu})\right)\ . \nonumber
\end{eqnarray}
\par
For the Wess-Zumino term we can write
\begin{eqnarray}
I_{WZ}&=&
\int d^6x\,
\epsilon^{\mu\nu\rho\sigma\tau\phi} \left[\frac{1}{6!}
c_{\mu\nu\rho\sigma\tau\phi}
\right.\nonumber\\ &&\left.
-\frac{1}{8}
\left(  \partial_{\mu} B_{\nu\rho }\right)
\bar \lambda\gamma_a\gamma_{a'}\partial_{\sigma}\lambda
\left( 2\delta_{\tau}^a-\bar \lambda\gamma^a\partial_\tau\lambda\right)
\partial_{\phi}X^{a'}\right]\ ,
\end{eqnarray}
where in the expression for $c_6$ in \eqn{c6} one has to replace $\theta$
by the chiral spinor $\lambda$, and can use \eqn{Pigf}.
\par
The action is invariant under the 32 global
supersymmetries on the worldvolume. The supersymmetry transformations for
the scalars $X^{a'}$ and the
spinors are given in \eqn{32susy} where again $A$ and $C$ have to be
taken at the values of $g_{\mu\nu}$ and ${\cal H}_{\mu\nu\rho}$
presented in \eqn{metric} and \eqn{3form}. The worldvolume
supersymmetry transformations of $B_{\mu\nu}$ can be derived
in analogous fashion from its $\kappa$-symmetry, space-time
supersymmetry and 6-dimensional diffeomorphism transformation.
Remark that also $a$ is not inert under these
transformations, due to $\delta a=\xi^\mu\partial_\mu a$. The
$SO(10,1)$ rotations of the original theory break up in linearly and
non-linearly realised symmetries. The linear ones are the
$SO(5,1)\times SO(5)$ transformations. The first factor gets from
$\xi^\mu$ in \eqn{decompxi} the extra contributions such that they
are the Lorentz rotations as in section~\ref{ss:conftr}. For the
$SO(5)$ transformations we have $\alpha^{a'b'}=-\Lambda^{a'b'}$.
The remaining
transformations in the coset $\frac{SO(10,1)}{SO(5,1)\times SO(5)}$
are non-linear in the gauge-fixed theory.
\subsection{Cutting to the linearized action}
By linearized, we imply that we count powers of $X^{a'}$, $\theta$
and $B_{\mu\nu}$, leaving $a$ arbitrary. We keep in the action terms
quadratic in these fields, and in transformation laws linear terms.
We obtain
\begin{eqnarray}
-{\cal G}&=&-1+\bar \lambda\dsl\lambda-\ft12 \partial_\mu
X^{a'}\partial^\mu X_{a'}+
\ft14 {\cal H}^*_{\mu\nu}{\cal H}^{*\mu\nu}\nonumber\\
c_3&=&0\nonumber\\
I_{WZ}&=&\int c_6=\int \ft1{5!}\epsilon^{\mu_1\ldots \mu_6}\bar \lambda
\gamma_{\mu_1\ldots \mu_5}
\partial_{\mu_6}\lambda=\int d^6x\,\bar \lambda\dsl\lambda
\end{eqnarray}
Putting everything together we have
\begin{equation}
S_{lin}= \int d^6 x \left[- \ft12 H^{*\mu\nu} H_{\mu\nu}^-
-\ft12 \partial_\mu X^{a'} \partial^\mu X_{a'} + 2 \bar \lambda \dsl
\lambda\right]\,.\label{slin}
\end{equation}
\par
For the transformation laws, the remaining parts of the
decomposition law are:
\begin{eqnarray}
\xi^\mu&=& a^\mu +\Lambda^\mu{}_\nu x^\nu  \nonumber\\
(1+C)\kappa&=&\ft12\gamma_{a'}\gamma_a
\Lambda^{a'a}\lambda-\epsilon\ .
\end{eqnarray}
The remaining part from the 6-dimensional general coordinate
transformations are thus the rigid 6-dimensional Poincar\'e
transformations (for the spinors they combine with the $6\times 6$
part of the 11-dimensional Lorentz transformations).
There remain also the translations in the extra 5 dimensions, as
constant shifts of the $X^{a'}$, and the Lorentz rotations in these 5
dimensions. These become the internal symmetry $USp(4)$.  The remaining
off-diagonal parts of the Lorentz
transformations remain only as
\begin{equation}
\delta_{off-d}X^{a'}=-\Lambda^{a'\mu}x_\mu\ .
\end{equation}
\par
Remains the supersymmetry part. We have
\begin{eqnarray}
\delta_f X^{a'}&=&-2\bar \epsilon_L\gamma^{a'}\lambda_R\nonumber\\
\delta_f \lambda_R&=&-A(1+C)^{-1}\epsilon_L\nonumber\\
\delta_f B_{\mu\nu}&=&-2\bar \epsilon_L\gamma_{\mu\nu}\lambda_R \ .
\end{eqnarray}
At the linear level we have
\begin{eqnarray}
\Gamma&=&\bar \gamma+\ft12\gamma^{\mu\nu\rho} {\cal H}^*_{\mu\nu}v_\rho
\nonumber\\
\bar \gamma&=& \pmatrix{1& \frac{1}{5!}\epsilon^{\mu_1\ldots
\mu_6}\Pi_{\mu_1}^{a'}\delta_{\mu_2\ldots \mu_6} ^{a_2\ldots a_6}
\gamma_{a'}\gamma_{a_2\ldots a_6}\cr -\frac{1}{5!}\epsilon^{\mu_1\ldots
\mu_6}\Pi_{\mu_1}^{a'}\delta_{\mu_2\ldots \mu_6} ^{a_2\ldots a_6}
\gamma_{a'}\gamma_{a_2\ldots a_6}&-1}
\nonumber\\
C&=&1\nonumber\\
A&=& -\gamma_{a'}\dsl X^{a'}  + \ft13
h^+_{\mu\nu\rho}\gamma^{\mu\nu\rho}\ .
\end{eqnarray}
For the last term, we use \eqn{chiralselfdual}, and
\begin{equation}
\left( {\cal H}^*_{[\mu\nu}v_{\rho]}\right) ^*=\ft43 v^\sigma
v_{[\sigma}{\cal H}_{\mu\nu\rho]}\ .
\end{equation}
Then we use the second line of \eqn{defh}.
This leads to the transformation
\begin{equation}
\delta_f \lambda_R=\ft12\gamma_{a'}\dsl X^{a'}\epsilon_L -
\ft16 h^+_{\mu\nu\rho}\gamma^{\mu\nu\rho}\epsilon_L\ .
\end{equation}
Thus we have derived the action for the small excitations of the M
5-brane from the  non-linear action. The supersymmetries
of the quadratic action follow from the supersymmetries of the full
non-linear action (however only 16 of the 32 are linear
supersymmetries acting on the states of the multiplet). The special
supersymmetries presented in Sec. 3. emerge only for quadratic
approximation
of the full M 5-brane theory.
\subsection{Breaking of superconformal symmetry by self
interaction of the tensor multiplet}
Consider the full non-linear action of the gauge-fixed M 5-brane
\eqn{SM5}. We can
expand it  around the quadratic approximation
\begin{equation}
S= {\rm const}  + \int d^6 x  ( -\ft12 H^*_{\mu\nu}H^{\mu\nu-}
- \ft12 (\partial X^{a'})^2+2\bar \lambda\dsl\lambda ) +
 S_{\rm interaction }
\ .
\end{equation}
Interaction terms are cubic, quartic etc. in scalars $X^{a'}$, spinors and
tensors. For example we have terms of the form
\begin{equation}
\partial_\mu X^{a'} \partial_\nu X^{a'} {\cal H}^{*\mu\nu}\ ,  \qquad
\partial_\mu X^{a'} \partial_\nu X^{a'} \bar \lambda \gamma^\mu
\partial^\nu
\lambda \ , \qquad  {\cal H}^{*\mu\nu}
{\cal H}_{\mu\nu} (\partial X^{a'})^2
\end{equation}
etc. All of these terms are not invariant under dilatation and therefore the
superconformal symmetry of the free action of the tensor multiplet is lost.
Another way to see this is to introduce the tension ${1\over l_P^6}$ and
rescale every field in the tensor multiplet by $l_P^3$. The quadratic part of
the action will be independent on $l_P$ however, the interaction terms will
all depend on $l_P$:
cubic terms will have  $l_P$ in front, quartic will have
 $l_P^2$ etc.
An interesting question which remains is to find out what will happen if the M
5-brane action would be quantized in the curved background of 11-dimensional
supergravity. This would correspond to taking into account terms proportional
to $l_P$ in the background. Is it possible to restore the superconformal
invariance of the free tensor multiplet in presence of both self-interaction
as well as interaction to supergravity? This we leave for further
investigations.

\section{$M5$ as a classical solution of 11-dimensional supergravity and
conformal symmetry}
One of the most puzzling issues of the current status of our
understanding of the fundamental theory is the relation between the
space-time and the worldvolume pictures. Here we suggest a step to
uncover this relation: we will study the symmetry of the space-time
near the horizon and symmetry of the quantum field theory on the
worldvolume.
\par
The space-time configuration corresponding to  M 5-brane is
\begin{equation}
ds^2=\left[1+\left( \frac{\mu}{\rho}\right) ^3
\right]^{-\ft13}(dx_\mu^2)+
\left[1+\left( \frac{\mu}{\rho}\right) ^3 \right] ^{\ft23}dx_{a'}^2\ ,
\end{equation}
where $\mu$ is a parameter and $dx_{a'}^2 =d\rho^2+\rho^2d^2\Omega$.
Near the horizon, $\rho\rightarrow 0$ \cite{GHT},
\begin{eqnarray}
ds^2 \ \ \ \rightarrow &&  \frac{\rho}{\mu} (dx_\mu^2)   +
\left( \frac{\mu}{\rho}\right) ^2 dx_{a'}^2   \nonumber\\
&&=\frac{\rho}{\mu} (dx_\mu^2)   + \left( \frac{\mu}{\rho}\right) ^2
d\rho^2+\mu^2d^2\Omega\ .
\end{eqnarray}
This is the $adS_7\times S^4$ geometry. We introduce the variable $w$
by
\begin{equation}
w=\left[1+\left( \frac{\mu}{\rho}\right) ^3 \right] ^{-\ft16}\ ,
\end{equation}
or solving for $\rho$:
\begin{equation}
\rho=\mu w ^2\left[1-w ^6 \right]
^{-\ft13}  =f(w)w^2\ ,
\end{equation}
where $f(w)$ is analytic at $w\rightarrow 0$. Thus
\begin{equation}
H\equiv  1+\left( \frac{\mu}{\rho}\right) ^3 = {1\over w ^{6}}\ .
\end{equation}
For $w\rightarrow 0$ the metric gets the $adS_7\times S^4$ form
\begin{equation}
ds^2 \rightarrow  \left[w^2dx_\mu^2+  (2\mu)^2 \left( \frac{dw}{w}
\right)
^2\right]+\mu^2 d^2\Omega\ .
\end{equation}
\par
The $adS_7$ transformations form the group $SO(6,2)$. This can be made
explicit by defining it as a submanifold of an 8-dimensional space
$$\{X^\mu,X^-,X^+\}$$
 with metric  (signature $(-++++++-)$)
\begin{equation}
ds^2=dX^\mu \eta_{\mu\nu} dX^\nu-dX^+dX^- \ .
\end{equation}
The submanifold is determined by the equation
\begin{equation}
X^\mu \eta_{\mu\nu} X^\nu-X^+X^- +(2\mu)^2 =0\ .
\end{equation}
On the hypersurface we will define the coordinates $\{x^\mu,w\}$ by
\begin{eqnarray}
X^-&=&w\nonumber\\
X^\mu &=& w x^\mu\nonumber\\
X^+&=&\frac{(2\mu)^2 +w^2x_\mu^2}{w}  \ .
\end{eqnarray}
The induced metric on the hypersurface is
\begin{equation}
ds^2=w^2 dx_\mu^2+\frac{(2\mu)^2 }{w^2}dw^2 \ .
\end{equation}
The $SO(6,2)$ is linearly realized in the embedding 8-dimensional
space, and these transformations, ($\hat\mu=\mu,+,-$ and
$\Lambda^{\hat\mu\hat\nu}=-\Lambda^{\hat\nu\hat\mu}$)
\begin{equation}
\delta X^{\hat\mu}=\Lambda^{\hat\nu\hat\rho}M_{\hat\nu\hat\rho}
X^{\hat\mu}=-\Lambda^{\hat\mu}{}_{\hat\nu} X^{\hat\nu}\ ,
\end{equation}
satisfy the algebra \eqn{algSO}. Some of these transformations are then
non-linearly realized on the $5+1$ space of the brane. Indeed we get
\begin{equation}
\delta x^\mu=-\xi^\mu\ ,
\end{equation}
with $\xi^\mu$ as in \eqn{ximu}, when we identify
\begin{eqnarray}
a^\mu=\Lambda^\mu{}_-+\frac{(2\mu)^2 }{w^2}\Lambda^\mu{}_+\ ;&\qquad &
\lambda_M^{\mu\nu}=\Lambda^{\mu\nu}\nonumber\\
\lambda_D=-\Lambda^-{}_-=\Lambda^+{}_+ \ ;&\qquad &
\Lambda_K^\mu=  \Lambda^\mu{}_+=\ft12\Lambda^{-\mu}\ .
\end{eqnarray}
\par
Thus we have found that the fact that the geometry of the infinite
throat is an
$adS_7$ space leads to the conformal symmetry of the 6-dimensional
worldvolume.
\section{Discussion}
We have performed here a study of the M 5-brane theory whose small excitations
are associated with  $OSp(8^*|2N)$ superconformal theory in d=6. Our main new
results are the following.
\par
We explained  the generic procedure which allows to establish the presence of
rigid superconformal symmetry in non-gravitational theories in dimensions
higher than 2.
\par
We have derived an action of the (0,2) free tensor multiplet in 6
dimensions.
The action has a superconformal symmetry and it is Lorentz
covariant due to the
use of the Pasti-Sorokin-Tonin \cite{auxil} auxiliary field $a(x)$.
The precise form of the superconformal invariant   action is
\begin{equation}
{\cal L}=
 -\ft12 H^*_{\mu\nu}H^{\mu\nu-}
- \ft12 (\partial X^{a'})^2+2\bar \lambda\dsl\lambda\ .
\end{equation}
where the two-index field $H_{\mu\nu} $ is a usual 3-index field
strength of
the tensor field contracted with the derivative of the auxiliary
field and
$*$
($-$) on $H$ mean dual (anti-self-dual) combinations. The Weyl weights and
properties
of fields under special conformal transformations, under
supersymmetry and
special supersymmetry and internal symmetry group can be found in
Sec. 4.
\par
This action was derived as a truncation of the full gauge-fixed M 5-brane
action. The meaning of scalar fields $X^{a'}(x) $ is that they are
excitation
of the five transverse directions of the brane in space-time, the
meaning of the
chiral
worldvolume spinors $\lambda(x) $ is that they are excitations of
(half) of
the 11-dimensional superspace coordinates $\theta(x) $ and finally,
$H(x)$
represent a tensor field on the brane which does not have a simple
interpretation in space time but is necessary to complete the
supersymmetric
tensor multiplet on the worldvolume. The supersymmetry
transformations of the
free tensor multiplet are
derived by the truncation of the non-linear supersymmetry, however
the special
supersymmetry emerges only for the quadratic action describing the small
excitations of the M 5-brane.
\par
The interaction terms of the M 5-brane action  violate
superconformal symmetry of the free action
since they have wrong scaling behavior, e. g. we have terms
$\partial_\mu X^{a'} \partial_\nu X^{a'} \bar \lambda \gamma^\mu
\partial^\nu
\lambda$ etc.
\par
We have given a  detailed derivation of the gauge-fixing procedure
for the full
non-linear M 5-theory with 32 linear-non-linear global
supersymmetries. This
leads to a better understanding of the single M 5-brane dynamics.
One may try to
generalize the gauge-fixing procedure for the case of the non-trivial
backgrounds. So far we have only worked in a flat 11-dimensional
superspace background.
\par
Finally we focused on the deep relation between $adS_7$ geometry of the
M 5-brane
throat and the supergroup generalization of it, which is precisely the
superconformal algebra of the small excitation of the brane. We explain this
relation in Sec. 6.
\par
We have
displayed the full superconformal algebra which corresponds to the
symmetry of the (0,2) tensor multiplet in six dimensions.
We have presented  the  $OSp(8^*|2N)$ algebra in two forms: first,
using a `triality' of $SO(6,
2)$ we gave a manifestly symmetric superalgebra with
graded (fermionic) bosonic (anti) commutators, see \eqn{superalgebra}.
Secondly, we presented
the $OSp(8^*|4)$ superconformal algebra in more familiar form which includes
28 generators of $SO(6,2)$ bosonic conformal symmetries and  16 supersymmetries
and 16 special supersymmetries and also
the generators of internal symmetry, an $USp(4)\approx {Spin(5)}$
group, see Sec. 3.
We have only studied here the classical superconformal algebra since
in d=6 even this is much less understood than in d=2 and in
d=4.\footnote{Jacques
 Distler  has informed us that he  also studied this  superconformal algebra
and its representations  (J. Distler, to appear).} The
next step will be to study the OPE's and extract information about
correlators of various operators in 6-dimensional superconformal
field theories.
\medskip
\section*{Acknowledgments.}
\noindent
We had stimulating discussions of various parts of the paper with M. Aganagic,
E. Bergshoeff,
J. Distler,
S. Ferrara, M. G{\" u}naydin, I. Klebanov,
J. Kumar, J. Maldacena, M. Peskin, J. Rahmfeld, A. Rajaraman, E. Silverstein,
J. H. Schwarz, S. Shenker, K. Stelle. The work of R. K. is
supported by the NSF grant PHY-9219345.
A.V.P. thanks the Physics Department of the Stanford University for
the hospitality during a fruitful visit in which this work was
performed.  He also thanks the FWO, Belgium, for the travel grant.
Work supported by
the European Commission TMR programme ERBFMRX-CT96-0045.
 \newpage
\appendix
\section{Notations}    \label{app:notations}
We start from 11-dimensional superspace with coordinates
\begin{equation}
Z^{\Lambda}=\{X^M,\theta^A\}\ .
\end{equation}
Let us recapitulate a list of indices and their range:
\begin{eqnarray}
M=0,1,\ldots ,10&& \mbox{11-dim. space-time}\nonumber\\
A=1,\ldots ,32 && 11-\mbox{dim. spinor indices, equivalent
with }\{(\alpha i),(\alpha' i)\}\nonumber\\
\alpha= 1,\ldots 4 && \mbox{chiral 6-dim. spinor indices}\nonumber\\
\alpha '= 1,\ldots 4 && \mbox{antichiral 6-dim. spinor
indices}\nonumber\\
i=1,\ldots ,4 && USp(4) \mbox{ indices}  \nonumber\\
\mu=0,1,\ldots ,5 && \mbox{curved on the 6-dimensional
worldsheet}\nonumber\\
a=0,1,\ldots ,5 && \mbox{flat on the 6-dimensional
worldsheet}\nonumber\\
a'=6,\ldots 10 && SO(5) \mbox{ indices}  \nonumber\\
\hat \mu=0,1,\ldots 7&&\mbox{(6,2) vector indices}  \nonumber\\
\hat\alpha=,1,\ldots 8&&\mbox{(6,2) chiral (right) spinor indices.}
\end{eqnarray}
The $X^M$ thus reduce to $\{X^\mu, X^{a'}\}$, the former being gauge
fixed to $x^\mu $, world-sheet coordinates on the 5-brane.
\par
The space-time metric $\eta_{MN}$ is $(-+\ldots +)$.
\par
Let us also repeat that all (anti)symmetrizations are with `weight
1', e.g. $A_{[\mu}B_{\nu]}=\ft12\left(  A_\mu B_\nu-A_\nu
B_\mu\right) $.
\subsection{Decomposing 11-dimensional gamma matrices
\ifnew \\ \else \newline \fi
and spinors}
\label{app:decomp11}
Many aspects about the Clifford algebras in various dimensions can
be found in \cite{KugoTown}. In 11 dimensions
the charge conjugation matrix is ${\cal C}$, which is antisymmetric,
and  ${\cal C}\Gamma_M$ is symmetric. Majorana spinors
satisfy $\bar \lambda\equiv -i\lambda^\dagger \Gamma_0=\lambda^T
{\cal C}$.
With indices, the matrix ${\cal C}$ is ${\cal C}^{AB}$, and
 ${\cal C}^{AB}{\cal C}_{CB}=\delta^A_C$, and gamma matrices are written as $\left(
\Gamma_M\right) _A{}^B$.
And let us also repeat the equations which are heavily used
\begin{eqnarray}
&& \Gamma^{MN}_{(AB}\Gamma_{CD)N}=0\nonumber\\
&&\Gamma_{M\,(AB}\Gamma^{MM_1M_2M_3M_4}_{CD)}=3\Gamma^{[M_1M_2}_{(AB}
\Gamma^{M_3M_4]}_{CD)}\ .
\end{eqnarray}
\par
We now decompose the 32-component spinor index $A$ in
$\{(\alpha i),(\alpha' i)\}$, and simultaneously $M$ in $\{a,a'\}$.
The $\alpha$ and $\alpha'$ will be chiral and antichiral indices in 6
dimensions as we will take a representation where
\begin{equation}
\Gamma_*\equiv -\Gamma_0\Gamma_1\Gamma_2\Gamma_3\Gamma_4\Gamma_5=
\pmatrix{\unity &0\cr 0&-\unity }\otimes \unity \ , \label{Gamma*}
\end{equation}
where in the notation with $\otimes $ the first factor refers to the
$8\times 8$ matrix in $(\alpha,\alpha')$, while the second factor
refers to the $4\times 4$ matrix in indices $i$.
In this sense we take a representation where
\begin{eqnarray}
{\cal C}&=&\pmatrix{0&c\cr c^T&0}\otimes \Omega  \nonumber\\
\Gamma_a&=&\pmatrix{0&\gamma_a\cr \tilde \gamma_a &0}\otimes \unity
\ ;\qquad
\Gamma_{a'}=\pmatrix{\unity &0\cr 0&-\unity }\otimes \gamma_{a'}\ ,
\label{calCGamma}\end{eqnarray}
where $\Omega$ is the (real antisymmetric) symplectic metric,
and there are further relations
\begin{eqnarray}
&&c\,c^\dagger =\unity \nonumber\\
&& \tilde\gamma_a \gamma_b + \tilde\gamma_b \gamma_a =2\eta_{ab}
\ ;\qquad
 \gamma_a \tilde\gamma_b + \gamma_b \tilde\gamma_a =2\eta_{ab}
\nonumber\\
&& \tilde\gamma_a^\dagger = \gamma_0\gamma_a\tilde\gamma_0 \
;\mbox{ or }
\left\{\begin{array}{l}
\tilde\gamma_a^\dagger= \gamma_a   \mbox{ for }a\neq 0   \\
\tilde\gamma_0^\dagger= -\gamma_0
\end{array} \right.    \nonumber\\
&& c^T\,\gamma_a=-\gamma_a^T\, c\ \mbox{implying also }
   c\,\tilde\gamma_a=-\tilde\gamma_a ^T\,c^T \nonumber\\
&&\gamma_{a'}
\gamma_{b'}+\gamma_{b'}\gamma_{a'}=2\delta_{a'b'}\nonumber\\
&&\gamma_{a'}^\dagger =\gamma_{a'} \ ;\qquad \trace \gamma_{a'}=0\ ;
\qquad\Omega \gamma_{a'} =-\left(\Omega \gamma_{a'}  \right) ^T\ .
\end{eqnarray}
We then have indeed symmetry of ${\cal C}\Gamma_M$:
\begin{equation}
{\cal C}\Gamma_ a= \pmatrix{c\,\tilde\gamma_a&0\cr 0&c^T\gamma_a}
\otimes \Omega\ ;\qquad {\cal C}\Gamma _{a'}=\pmatrix{0&-c\cr c^T&0}
\otimes \Omega \gamma_{a'}\ .
\end{equation}
Finally \eqn{Gamma*} imposes on our representation that
\begin{equation}
\gamma_0\tilde\gamma_1\gamma_2\tilde\gamma_3\gamma_4\tilde\gamma_5=
-\unity \ ,
\end{equation}
which then also implies $
\tilde\gamma_0\gamma_1\tilde\gamma_2\gamma_3\tilde\gamma_4\gamma_5=
\unity $.
\par
To indicate our use of indices, we rewrite some matrices from above with
indices:
\begin{eqnarray}
&&{\cal C}^{AB}=\pmatrix{0&c^{\alpha\beta'}\Omega^{ij}\cr
c^{\alpha'\beta}\Omega^{ij}&0} 
\ ;\qquad
{\cal C}_{AB}=\pmatrix{0&c_{\alpha\beta'}\Omega_{ij}\cr
c_{\alpha'\beta}\Omega_{ij}&0} \nonumber\\
&&\left( \Gamma_a\right)_A{}^B=
\pmatrix{0&\left( \gamma_a\right) _\alpha{}^{\beta'}\delta_i{}^j\cr
\left( \gamma_a\right) _{\alpha'}{}^\beta\delta_i{}^j&0}\nonumber\\
&&\left( \Gamma_a\right) ^{AB}\equiv
{\cal C}^{AC}\left( \Gamma_a\right)
_C{}^B=\pmatrix{\gamma_a^{\alpha\beta}
\Omega^{ij} &0\cr 0&\gamma_a^{\alpha'\beta'}\Omega^{ij}} \nonumber\\
&&\left( \Gamma_{a'}\right)_A{}^B=\pmatrix{\delta_\alpha{}^{\beta}
\left( \gamma_{a'}\right) _i{}^j&0\cr 0&-\delta_{\alpha'}{}^{\beta'}
\left( \gamma_{a'}\right) _i{}^j}\ ,
\end{eqnarray}
where
\begin{equation}
c^{\alpha\alpha'} =c^{\alpha'\alpha}
=\left( c_{\alpha\alpha'}\right) ^*  \ ;\qquad
c^{\alpha\alpha'} c_{\alpha'\beta}=\delta^\alpha{}_{\beta}\ :\qquad
\Omega_{ij}=-(\Omega^{ij})^{-1}\ .
\end{equation}
The matrix $\tilde\gamma_a$ is now
the matrix $\gamma_a$ where the first (lower) index is the antichiral
$\alpha'$, and the second is $\alpha$. We then have (for $a\neq 0$)
\begin{equation}
\left( \left( \gamma_a\right) _{\alpha'}{}^\alpha  \right) ^*=
\left( \gamma_a\right) _{\alpha}{}^{\alpha'}\ ,
\end{equation}
and the indices of this matrix are raised or lowered with
$c^{\alpha\alpha'}$ using the NW-SE convention:
\begin{equation}
\left( \gamma_a\right) ^{\alpha\beta}=c^{\alpha\alpha'}
\left( \gamma_a\right) _{\alpha'}{}^{\beta}\ ;\qquad
\left( \gamma_a\right) _{\alpha\beta}=
\left( \gamma_a\right) _{\alpha}{}^{\alpha'}c_{\alpha'\beta}\ .
\end{equation}
Remark that the Majorana condition is
\begin{equation}
\bar \lambda=\lambda^T{\cal C}\qquad\mbox{or}\qquad
\bar \lambda^A=\lambda_B{\cal C}^{BA}=-{\cal C}^{AB}\lambda_B=-\lambda^A\ ,
\end{equation}
using again the NW-SE index contraction.
\subsection{6-dimensional conventions}
Writing the latter equation on the spinor split in 6-dimensional
chiral parts,
$\lambda=\pmatrix{\lambda_{\alpha\,i}\cr \lambda_{\alpha'\, i}}$, we
obtain the 6-dimensional Majorana condition which we will adopt
\begin{equation}
\bar \lambda^{i\alpha'}\equiv -i\left( \lambda_{i\beta}\right) ^\dagger
\left( \gamma_0\right) _\beta{}^{\alpha'}=\lambda_{j\beta}\Omega^{ji}
c^{\beta\alpha'}=-\lambda^{i\alpha'} \ ,    \label{Majd6}
\end{equation}
which also applies for indices $\alpha$ and $\alpha'$ interchanged,
and where the last equation uses raising of indices with NW-SE
convention:
\begin{equation}
\lambda_i=\lambda^j \Omega_{ji}\ ;\qquad \lambda^i=
\Omega^{ij}\lambda_j\ .
\end{equation}
\par
We call a spinor of the form
\begin{equation}
\lambda=\pmatrix{\lambda_{\alpha i}\cr 0}\qquad\mbox{"left handed" or
"positively chiral"} \ .
\end{equation}
They thus satisfy $\lambda=\Gamma_*\lambda$.
Note that the Majorana conjugate has the $\alpha'$ index (up).
\par
The charge conjugation matrix in 6 dimensions is $c^{\alpha\alpha'}$,
and the gamma matrices are the first factor of $\Gamma_a$ in
\eqn{calCGamma}. Writing  $L$ and $R$ for left- and
right-handed spinors we have
\begin{eqnarray}
\bar \lambda_L\Gamma^a\theta_L= \bar \lambda_L\gamma^a\theta_L \
;&\qquad&
\bar \lambda_R\Gamma^a\theta_R= \bar \lambda_R\gamma^a\theta_R
\nonumber\\
\bar \lambda_R\Gamma^{a'}\theta_L= \bar \lambda_R\gamma^{a'}\theta_L \
;&\qquad&
\bar \lambda_L\Gamma^{a'}\theta_R= -\bar
\lambda_L\gamma^{a'}\theta_R\ ,
\end{eqnarray}
such that although $\Gamma^a$ and $\Gamma^{a'}$ anticommute, $\gamma^a$ and
$\gamma^{a'}$ commute.
\par
We define
\begin{equation}
\epsilon_{012345}=1=-\epsilon^{012345}\ ,
\end{equation}
and we thus have
\begin{equation}
\epsilon^{a_1\dots a_m\, b_1\dots b_{6-m}} \epsilon_{a_1\dots a_m\,
c_1\dots
c_{6-m}} = - (6-m)! m! \delta_{[c_1}^{b_1} \dots
\delta_{c_{6-m]}}^{b_{6-m}}\ .
\end{equation}
It is useful to know the relation
\begin{equation}
\gamma^{a_1\ldots a_k}=\frac{S_k}{(6-k)!}\epsilon^{a_1\ldots a_6}
\gamma_{a_{k+1}\ldots a_6}\Gamma_*\ ,  \qquad S_k=\left\{
\begin{array}{l}
+1:\quad k=0,1,4,5\\
-1:\quad k=2,3,6
\end{array}     \right.
\end{equation}
where $\Gamma_*$  is $+1$ on left-handed and $-1$ on right-handed
spinors. We denote the dual, self-dual and anti-self-dual of a three
index tensor as
\begin{equation}
F^* _{abc}= \ft16 \epsilon_{abcdef} F^{def}\ ;\qquad
F_{abc}^\pm=\ft12\left( F_{abc}\pm F_{abc}^* \right)\ .\label{defdual}
\end{equation}
\par
Remark that omitting the $\Omega^{ij}$ in ${\cal C}$ this becomes the
symmetric charge conjugation matrix from 6 dimension.
\par
As we did already when using explicit indices in the second part of
appendix~\ref{app:decomp11}, we will omit tildes on the $\gamma$
matrices when chiral spinors are used, as the place where these
matrices occur shows automatically which one is meant.
\par
For a left-handed spinor $\epsilon$ we have for any antisymmetric
tensor $F_{abc}$:
\begin{equation}
F_{abc}\gamma^{abc}\epsilon= F^+_{abc}\gamma^{abc}\epsilon
 \ ;\qquad
\gamma_d F^+_{abc}\gamma^{abc}\epsilon= 6
F^+_{abd}\gamma^{ab}\epsilon\ .
\label{chiralselfdual}
\end{equation}
\par
Under complex conjugation we have
\begin{equation}
\left( \bar \lambda \chi\right) ^*=
\left( \bar \lambda^i \chi_i\right) ^*=
\chi_i^\dagger \left( \bar \lambda^i\right) ^\dagger=
\bar \chi^i \lambda_i =\bar \lambda^i\chi_i=
\bar \chi \lambda =
\bar \lambda\chi\ .
\end{equation}
\subsection{$SO(5)$ notation}\label{app:SO5}
We can replace $USp(4)$ notation with $SO(5)$. We identify
$\Omega_{ij}$ with the charge conjugation matrix in 5 dimensions,
which is thus real and antisymmetric.
An $\Omega$-traceless antisymmetric tensor $X^{ij}$ can then be
translated to a 5-dimensional vector, and a symmetric tensor $U_{ij}$
to an antisymmetric $U_{a'b'}$:
\begin{eqnarray}
X^{a'}= \ft12\gamma^{a'}_{ij} X^{ij}\ ;&\qquad &
X^{ij} =\ft12 \gamma^{a'ij} X_{a'}\nonumber\\
U_{a'b'}=\ft14 \gamma_{a'b'}^{ij}U_{ij}\ ;&\qquad &
U_{ij}=\ft12 \gamma_{ij}^{a'b'} U_{a'b'} \ .
\end{eqnarray}
Note that this implies
\begin{equation}
\delta_U\equiv \ft12\alpha^{ij}U_{ij} =
\ft14 \alpha^{ij}\gamma_{ij}^{a'b'} U_{a'b'}=
\alpha^{a'b'} U_{a'b'} \ .
\end{equation}
For vectors and spinors of $Spin(5)$ we have
\begin{equation}   \delta_U X^{a'} = \alpha^{a'b'}X_{b'}\ ;\qquad
\delta_U \lambda   = \ft14 \alpha^{c'd'}\gamma_{c'd'}\lambda \ .
\end{equation}
\par
We will often therefore omit the $USp(4)$ indices, implying a NW-SE
convention for summation indices:
\begin{equation}
\bar \lambda \xi = \bar \lambda^i \xi_i\ ;\qquad
\bar \lambda \gamma_{a'} \xi = \bar \lambda^i
(\gamma_{a'})_i{}^j\xi_j\ ,
\end{equation}
so $\gamma_{a'}$ matrices are supposed to have their indices in the
position $(\gamma_{a'})_i{}^j$.
\par
When taking Majorana conjugates, $\gamma_a$, $\gamma_{ab}$ and
$\gamma_{a'b'}$ are the matrices which lead to minus signs.
E.g. in a supersymmetry commutator we
can only have the following structures:
\begin{equation}
\bar \epsilon_1  \gamma_a \epsilon_2\ ;\qquad
\bar \epsilon_1  \gamma_a \gamma^{a'}\epsilon_2\ ;\qquad
\bar \epsilon_1  \gamma_{abc} \gamma^{a'b'}\epsilon_2\ .
\end{equation}
This gives thus the antisymmetric structures between spinors of equal
chirality. For spinors of opposite chirality the symmetric
combinations are
\begin{equation}
\bar \lambda\xi=\bar \xi\lambda\ ;\qquad
\bar \lambda\gamma^{a'}\xi=\bar \xi\gamma^{a'}\lambda\ ;\qquad
\bar \lambda\gamma^{a'b'}\gamma_{ab}\xi=
\bar \xi\gamma^{a'b'}\gamma_{ab}\lambda\ .
\end{equation}
\subsection{$(6,2)$ Clifford algebra}  \label{app:Cliff62}
The general properties of this algebra can once more be extracted
from \cite{KugoTown}.    We
start from the 6-dimensional matrices in \eqn{calCGamma} (omitting
the $USp(4)$ part). The $SO(6,2)$ indices will be indicated by $\hat
\mu=0,1,\ldots ,7$, and we use as signature $(-++++++-)$. The
gamma-matrices and charge conjugations are $16\times 16$
matrices\footnote{We work here in flat space where $\gamma_\mu$ are
the constant matrices $\gamma_a$ from above.}
\begin{eqnarray}
\hat \Gamma_{\mu}&=&\pmatrix{0&\gamma_\mu\cr \tilde \gamma_\mu &0}\otimes
\sigma_3 
\ ;\qquad
\hat \Gamma_6=\unity \otimes \sigma_1 \ ;\qquad
\hat \Gamma_7=\unity \otimes (-i)\sigma_2\nonumber\\
\hat {\cal C}&=&\pmatrix{0&c\cr c^T&0}\otimes \sigma_1= \hat {\cal C}^T
\ ;\qquad
\left(\hat {\cal C}\hat \Gamma_{\mu} \right) ^T=
\hat {\cal C}\hat \Gamma_{\mu}   \label{gammad8} \ .
\end{eqnarray}
The second factor here is thus not in the internal $USp(4)$ space as in
\eqn{calCGamma},
but are $2\times 2$ matrices, with $\sigma_1\sigma_2=i\sigma_3$. To
define chirality we first obtain
\begin{equation}
\hat \Gamma_*=-\hat \Gamma_ 0\hat \Gamma_1\ldots \hat \Gamma_7=
\pmatrix{\unity &0\cr 0&-\unity }\otimes \sigma_3\ .
\end{equation}
We will indicate with $\hat \alpha$ the index of right-handed spinors
in 8 dimensions. They are composed of a
right-handed and a left-handed one in 6-dimensions as follows
\begin{equation}
\hat \lambda_{\hat \alpha}= \left( \pmatrix{0 \cr\lambda_{\alpha'}}\otimes
\pmatrix{1\cr 0} \right) + \left(\pmatrix{\lambda_{\alpha}\cr 0}\otimes
\pmatrix{0\cr 1}  \right) \ .      \label{chiral62spinor}
\end{equation}
Its two parts are separated as eigenvectors of
\begin{equation}
\hat \Gamma_{67}=\unity \otimes \sigma_3\ . \label{Gamma67}
\end{equation}
Further, we have in the chiral subspace, where
\begin{equation}
\hat \lambda_{\hat \alpha}= \pmatrix{\lambda_{\alpha'}\cr\lambda_{\alpha}}
\ ,    \label{chiral62spinor2}
\end{equation}
that
\begin{eqnarray}
\hat {\cal C}=\pmatrix{0&c^T\cr c&0}\ ;&\qquad&
\hat \Gamma_{ \mu \nu}= \pmatrix{\tilde\gamma_{[\mu}\gamma_{\nu]}
&0\cr 0&\gamma_{[\mu}\tilde\gamma_{\nu]}} \nonumber\\
\hat \Gamma_{\mu 7}+\hat \Gamma_{\mu 6}=
\pmatrix{0&0\cr -2 \gamma_\mu &0} \ ;&\qquad&
\hat \Gamma_{\mu 7}-\hat \Gamma_{\mu 6}=
\pmatrix{0&-2\tilde\gamma_\mu\cr 0 &0} \ . \label{Gamma622}
\end{eqnarray}
Let us repeat that we drop the tildes in the main text where these
matrices act on chiral or antichiral $6d$ spinors.
\subsection{Forms and integration}\label{app:forms}
We define the tangent space components of a generic
$p$--forms, $\phi_p$, according to
\begin{equation}
\label{dec}
\phi_p={1\over p!}e^{\mu _1}\cdots e^{\mu _p}\phi_{\mu _p \cdots
\mu _1},
\end{equation}
where the wedge product between forms will always be understood.
The following relations are easily derived,
\begin{equation}
dx^{\mu_0}\,\dots\, dx^{\mu_5}=-\epsilon^{\mu_0\dots\mu_5}
dx^0\,\dots \, dx^5= -\epsilon^{\mu_0\dots\mu_5} d^6 x\ .
\end{equation}
It follows that
\begin{equation}
\int e^\mu e^\nu e^\rho e^\sigma e^\tau e^\phi =\int d^6 x\,
\epsilon^{\phi\tau\sigma\rho\nu\mu}
=-\int d^6 x\,\epsilon^{\mu\nu\rho\sigma\tau\phi}\ ,
\end{equation}
and for the 3-forms $H$ and $G$
\begin{equation}
\int G\, H = \int e^\mu e^\nu e^\rho e^\sigma e^\tau e^\phi
\ft16\ft16 G_{\mu\nu\rho} H_{\sigma\tau\phi} = -\ft16
\int d^6x \sqrt g G_{\mu\nu\rho} H^{*\,\mu\nu\rho}\ .
\end{equation}
Note also that e.g. in contrast to \cite{ag} we consider the
differentials as space-time derivatives, commuting with the spinors.
\section{More on the conformal algebra}      \label{app:conf}
In the general rule for the transformation of a field, \eqn{deltaC},
appears the Lorentz transformation matrix $m_{\mu\nu}$, which should
satisfy
\begin{equation}
m_{\mu\nu}{}^i{}_k m_{\rho\sigma}{}^k{}_j  -
m_{\rho\sigma}{}^i{}_k m_{\mu\nu}{}^k{}_j =
-\eta_{\mu[\rho}m_{\sigma]\nu}{} ^i{}_j
+\eta_{\nu[\rho}m_{\sigma]\mu} {}^i{}_j \ .
\end{equation}
Note the sign difference between the commutator of these matrices and
the commutator of the generators, which is due to the difference
between `active and passive' transformations.  See e.g. also for
transformations on a field of zero Weyl weight:
(transformations act only on fields, not on explicit $x^\mu$)
\begin{eqnarray}
\lambda_D a^\mu [D,P_\mu]\phi(x)&=& \left( \delta_D(\lambda_D)
\delta_P(a^\mu)-\delta_P(a^\mu) \delta_D(\lambda_D)\right) \phi(x)
\nonumber\\ &=&
\delta_D(\lambda_D)  a^\mu \partial_\mu \phi(x)   - \delta_P(a^\mu)
\lambda_D x^\mu \partial_\mu \phi(x) \nonumber\\ &=&
a^\mu \partial_\mu  \left( \lambda_D x^\nu \right)  \partial_\nu
\phi(x)
\nonumber\\
&=& a^\mu \lambda_D \partial_\mu  \phi(x)  = \lambda_D a^\mu
P_\mu\phi(x) \ .
\end{eqnarray}
The explicit form for Lorentz transformation matrices is for
vectors (the indices $i$ and $j$ are of the same kind as $\mu$
and $\nu$)
\begin{equation}
m_{\mu\nu}{}^\rho{}_\sigma =-\delta^\rho_{[\mu}\eta_{\nu]\sigma} \ ,
\end{equation}
while for spinors, (where $i$ and $j$ are (unwritten) spinor indices)
\begin{equation}
m_{\mu\nu} =-\ft14\gamma_{\mu\nu}\ .
\end{equation}
\par
It is interesting to see how the transformations of derivatives of
fields get similar to those of covariant derivatives in gauged
conformal gravity. E.g. for a scalar of weight $w$ (and without
extra special conformal transformations) we get
\begin{eqnarray}
\delta_C  \partial_\mu\phi(x)&=&\xi^\nu(x)\partial_\nu\partial_\mu \phi(x)
+w\,\Lambda_D(x)\,\partial_\mu\phi(x)\nonumber\\ &&
-\Lambda_{M\mu}{}^\nu (x)\partial_\nu \phi(x) +\Lambda_D(x)
\partial_\mu \phi(x) -2w\,\Lambda_{K\mu}  \phi(x) \ .  \label{delCderphi}
\end{eqnarray}
If one would consider the covariant derivatives in a local approach
then the appropriate covariant derivative is
\begin{equation}
D_a \phi(x) = e_a^\mu \left(\partial_\mu-w\,b_\mu\right)\phi(x)\ .
\end{equation}
Due to a `theorem on covariant derivatives' (see section 2.5 of
\cite{Karpacz2}) one only has to consider some transformations of the
gauge fields to obtain the full result of its transformation. One
discovers that the last line of \eqn{delCderphi} exactly contains
these extra terms.

\section{Details on M 5-brane and $\kappa$-symmetry} \label{app:kappa}
To determine the $\kappa$ transformations, we first do not fix
$\dkt$. The transformation of $X^M$ in \eqn{transformations} is
sufficient to determine that $c_3$ transforms as
\begin{equation}
\delta_\kappa c_3=(\delta_\kappa\bar \theta
)\Gamma_{MN}d\theta\,\Pi^M\Pi^N +d(\mbox{something})\ .
\end{equation}
We then fix the $\kappa$-transformation of $B$ such that it cancels
with the second term above in $\dk{\cal H}$. We thus obtain $\dk B$
as in \eqn{deltakapparem} and
\begin{eqnarray}
\delta_\kappa c_3&=&(\delta_\kappa\bar \theta
)\Gamma_{MN}d\theta\,\Pi^M\Pi^N + \ft12d \delta_\kappa B \nonumber\\
\delta_\kappa {\cal H}&=&\Pi^{M_1}\Pi^{M_2} d\bar \theta\Gamma_{M_1M_2}
\delta_\kappa \theta\quad\mbox{or}\quad
\delta_\kappa {\cal H}_{\mu\nu\rho} = 6 (\delta_\kappa \bar \theta)
\gamma_{[\mu\nu} \partial_{\rho]}\theta\ .
\end{eqnarray}
This fixes all the transformations in terms of $\dkt$.
\par
Now calculate progressively the $\kappa$-transformations of the
objects which appear in the action:
\begin{eqnarray}
\dk g_{\mu\nu}&=&
4\left( \partial_{(\mu}\bar \theta\right) \gamma_{\nu)}\dkt
\ ;\qquad
\dk \sqrt{g}=2\sqrt{g} \left( \partial_{\mu}\bar \theta\right)
\gamma^\mu \dkt\nonumber\\
\dk R_7&=&\frac{2}{4!}d\bar \theta \Gamma^{M_1}\dkt \, d\bar \theta
\Gamma_{M_1\ldots M_5}d\theta \Pi^{M_1}\cdots \Pi^{M_4}\nonumber\\
&&+\frac{2}{5!}d\bar \theta
\Gamma_{M_1\ldots M_5}d\dkt \Pi^{M_1}\cdots \Pi^{M_5}\nonumber\\
\dk R_4&=&2d\bar \theta \Gamma^{M_1}\dkt \, d\bar \theta
\Gamma_{M_1 M_2}d\theta \Pi^{M_2}+d\bar \theta
\Gamma_{M_1 M_2}d\dkt \Pi^{M_1}\Pi^{M_2}\nonumber\\
\dk I_7&=&d\left( -\frac{2}{5!}d\bar \theta
\Gamma_{M_1\ldots M_5}\dkt \Pi^{M_1}\cdots \Pi^{M_5}  -\frac12 {\cal
H} d\bar \theta \Gamma_{M_1 M_2}\dkt
\Pi^{M_1}\Pi^{M_2}\right)\nonumber \\
\dk \sqrt{u^2}&=&-\ft12\sqrt{u^2}  v^\mu v^\nu \dk g_{\mu\nu}\qquad \rightarrow
\dk v_\mu=\ft12 v_\mu v^\nu v^\rho \dk g_{\nu\rho}\nonumber\\
\dk{\cal H}^{*\mu\nu}&=&-\frac{1}{\sqrt{g}}
\epsilon^{\mu\nu\rho\sigma\tau\phi}
v_\rho \partial_\sigma\theta\gamma_{\tau\phi}\dkt   \nonumber\\
&& +{\cal H}^{*\mu\nu} \left( -2
 \left( \partial_{\rho}\bar \theta\right) \gamma^\rho \dkt
+\ft12    v^\rho v^\sigma\dk g_{\rho\sigma}\right)  \nonumber\\
\dk {\cal H}_{\mu\nu}&=&\left( \dk {\cal H}_{\mu\nu\rho}\right) v^\rho
- {\cal H}_{\mu\nu}\frac{ \dk \sqrt{u^2}}{ \sqrt{u^2}}+
{\cal H}_{\mu\nu\rho}v_\sigma \dk g^{\rho\sigma}\ .
\end{eqnarray}
We define  (inspired by \cite{ag}, but with some other factors)
\begin{eqnarray}
&&\dk\left( -\int d^6 x \frac{\sqrt{g}}4 {\cal H}^{*\mu\nu}
{\cal H}_{\mu\nu} +I_{WZ}  \right) =
\frac{\sqrt{g}}{2}\partial_\mu\bar \theta   T^\mu\dkt
\nonumber\\
&&\dk{\cal G}=\frac{g}{2{\cal G}} \partial_\mu\bar \theta
U^\mu\dkt\ .
\end{eqnarray}
We make use of the expression $\bar \gamma$, defined in \eqn{resGamma},
and satisfying
\begin{equation}
\bar \gamma^2=1\ ;\qquad
\gamma^{\mu_{k+1}\ldots \mu_6}\bar
\gamma=(-)^{(k+1)(k+2)/2}\frac{1}{k!\sqrt{g}}\epsilon^{\mu_1\ldots
\mu_6}
\gamma_{\mu_1\ldots \mu_k}\ .
\end{equation}
This leads to
\begin{eqnarray}
T^\mu&=& 4\gamma^\mu\bar \gamma+6\gamma_{\nu\rho}{\cal H}^{*[\mu\nu}
v^{\rho]}+2{\cal H}^*_{\nu\rho} {\cal
H}^{\nu\rho(\mu}v^{\sigma)}   \gamma_\sigma
-2{\cal H}^{*\nu\rho} {\cal H}_{\nu\rho} v^\mu
v^\sigma\gamma_\sigma \nonumber\\
&=&4\gamma ^\mu \bar \gamma  +6\gamma_{\nu\rho}{\cal
H}^{*[\mu\nu} v^{\rho]} - \ft1{\sqrt g}
\epsilon^{\rho\sigma\lambda\tau\phi(\mu}
v^{\nu)} {\cal H}^*_{\rho\sigma} {\cal
H}_{\lambda\tau}^* v_\phi\gamma_\nu\ ,
\end{eqnarray}
using
\begin{equation}
{\cal H}_{\mu\nu\rho} = 3 v_{[\mu}{\cal H}_{\nu\rho]} + \ft12{\sqrt g}
\epsilon_{\mu\nu\rho\sigma\lambda\tau}{\cal
H}^{*\sigma\lambda} v^\tau\ .
\end{equation}
\par
To calculate $U_\mu$, we make use of \eqn{BI}:
\begin{eqnarray}
\delta_\kappa {\cal G} &=& \frac{1}{2{\cal G}} \delta_\kappa {\cal
G}^2\nonumber\\
\ft1g\delta_\kappa {\cal G}^2&=&  g^{\mu\nu} \dk g_{\mu\nu} \left(1 - \ft12
{\cal H}^*_{\rho\sigma}{\cal H}^{*\rho\sigma} - \ft38
{\cal H}^{*\lambda\tau}{\cal H}_{\tau[\rho}^*{\cal H}^*_{\sigma\lambda]}{\cal
H}^{*\rho\sigma}\right)\nonumber\\
&& - \left({\cal H}^*_{\rho\sigma} + \ft32 {\cal H}^{*\lambda\tau}
{\cal H}_{\tau[\rho}^*{\cal H}_{\sigma\lambda]}^*\right) \dk {\cal
H}^{*\rho\sigma}\nonumber\\
&=& g^{\mu\nu} \dk g_{\mu\nu} \left(1 - \ft12
{\cal H}^*_{\rho\sigma}{\cal H}^{*\rho\sigma} - \ft38
{\cal H}^{*\lambda\tau}{\cal H}_{\tau[\rho}^*{\cal H}^*_{\sigma\lambda]}{\cal
H}^{*\rho\sigma}\right)\nonumber\\
&&-{\cal
H}^*_{\rho\sigma}\left[-\ft1{\sqrt{g}}\epsilon^{\rho\sigma\lambda\tau\mu\nu}
v_\lambda\partial_\tau\bar
\theta \gamma_{\mu\nu} \dkt - \ft12 {\cal H}^{*\rho\sigma} g^{\mu\nu} \dk
g_{\mu\nu}\right.
\nonumber\\
&&\hskip2.5cm \left.+2{\cal H}^{*\rho\sigma} v^\mu v^\nu \partial_\mu\bar\theta
\gamma_\nu \dkt -
4 {\cal H}^{*\mu \rho} \partial_\mu \bar\theta \gamma^\sigma \dkt\right]
\nonumber\\
&&-\ft32 {\cal H}^{*\lambda\tau}
{\cal H}_{\tau[\rho}^*{\cal
H}_{\sigma\lambda]}^*\left[-\ft1{\sqrt g}
\epsilon^{\rho\sigma\mu\nu\xi\phi}v_\mu\partial_\nu\bar
\theta \gamma_{\xi\phi} \dkt - \ft12 {\cal H}^{*\rho\sigma} g^{\mu\nu} \dk
g_{\mu\nu}\right.
\nonumber\\
&&\hskip2.5cm \left.+2{\cal H}^{*\rho\sigma} v^\mu v^\nu \partial_\mu\bar\theta
\gamma_\nu \dkt
- 4 {\cal H}^{*\mu \rho} \partial_\mu \bar\theta \gamma^\sigma \dkt\right]\ .
\end{eqnarray}
This leads to
\begin{eqnarray}
U^\mu&=&4\gamma^\mu-\ft1{\sqrt g} \epsilon^{\mu\nu\rho\lambda\sigma\tau}{\cal
H}^*_{\nu\rho}v_\lambda
\gamma_{\sigma\tau} - 4{\cal H}^*_{\nu\rho}{\cal H}^{*\mu\rho}\gamma^\nu
-2 {\cal H}^*_{\rho\sigma} {\cal H}^{*\rho\sigma} v^\mu v^\nu
\gamma_\nu\nonumber\\
&&-\ft32{\cal H}^{*\tau\phi}{\cal H}^*_{\phi[\nu} {\cal H}^*_{\rho\tau]}
\left[\ft1{\sqrt g} \epsilon^{\mu\nu\rho\lambda\xi\phi}v_\lambda
\gamma_{\xi\phi}-{\cal
H}^{*\nu\rho}\gamma^\mu + 4{\cal H}^{*\mu\rho}\gamma^\nu \right.\nonumber\\
&&\left.\hskip2.5cm + 2 {\cal H}^{*\nu\rho} v^\mu v^\lambda
\gamma_\lambda \right]\ .
\end{eqnarray}
The quantities $T^\mu$ and $U^\mu$ are related by
\begin{equation}
U^\mu = T^\mu \rho\ \mbox{ with}\ \rho=
\bar \gamma + \ft{1}2{\cal H}_{\mu\nu}^*v_\rho
\gamma^{\mu\nu\rho} - \ft1{16\sqrt{g}}\epsilon^{\mu\nu\rho\sigma\tau\phi}
{\cal H}^*_{\mu\nu} {\cal H}^*_{\rho\sigma} \gamma_{\tau\phi} \ .
\label{UisTrho}
\end{equation}
The quantity $\rho$ squares to
\begin{equation}
\rho^2=\frac{{\cal G}^2}{g}\ .    \label{rho2}
\end{equation}
\par
The kappa invariance thus requires
\begin{equation}
\frac{\sqrt g}{2} \partial_\mu \bar\theta T^\mu
\left( 1 -\frac{\sqrt{g}}{{\cal G}}\rho\right )\dkt=0\ .
\end{equation}
This leads us to define
\begin{equation}
\Gamma =  \frac{\sqrt g}{{\cal G}} \rho \ ,       \label{resGamma2}
\end{equation}
consistent with $\Gamma^2=1$. Indeed, then invariance is obtained by
the rule \eqn{dktrule}, and $\Gamma$ is given by \eqn{resGamma}.

\end{document}